\documentclass[preprintnumbers,article,amsmath,amssymb,floatfix,10pt,prd,onecolumn,superscriptaddress,nofootinbib]{revtex4-2}

\usepackage{titlesec}
\usepackage[toc]{appendix}
\usepackage{caption}
\titleformat*{\section}{\LARGE\bfseries}
\titleformat*{\subsection}{\Large\bfseries}
\titleformat*{\subsubsection}{\large\bfseries}
\titleformat*{\paragraph}{\large\bfseries}
\titleformat*{\subparagraph}{\large\bfseries}
\usepackage{bm}
\usepackage{amsfonts}
\usepackage{latexsym}
\usepackage[utf8]{inputenc}
\usepackage{graphicx}
\usepackage{amsmath}
\usepackage{palatino}
\usepackage{mathpazo}
\usepackage{textcomp}
\usepackage{float}
\usepackage{booktabs}
\usepackage{dcolumn}
\usepackage{ragged2e}
\usepackage{hyperref}
\hypersetup{colorlinks,citecolor=blue}
\hypersetup{colorlinks=true,linkcolor=blue,filecolor=magenta,urlcolor=blue}
\usepackage{amsthm}
\usepackage{xcolor}
\usepackage{orcidlink}
\usepackage{epsfig}
\usepackage{commath}
\usepackage{cancel, ulem}
\usepackage{subcaption}
\usepackage{caption}
\usepackage{multirow}
\newcommand{\m}{\mathring}
\newcommand{\be}{\begin{equation}}
\newcommand{\ee}{\end{equation}}
\newcommand{\bea}{\begin{eqnarray}}
\newcommand{\eea}{\end{eqnarray}}
\newcommand{\eeas}{\end{eqnarray*}}
\newcommand{\beas}{\begin{eqnarray*}}

\def\jnl@style{\it}
\def\aaref@jnl#1{{\jnl@style#1}}

\def\aaref@jnl#1{{\jnl@style#1}}

\def\aj{\aaref@jnl{AJ}}                   
\def\apj{\aaref@jnl{ApJ}}                 
\def\apjl{\aaref@jnl{ApJ}}                
\def\apjs{\aaref@jnl{ApJS}}               
\def\apss{\aaref@jnl{Ap\&SS}}             
\def\aap{\aaref@jnl{A\&A}}                
\def\aapr{\aaref@jnl{A\&A~Rev.}}          
\def\aaps{\aaref@jnl{A\&AS}}              
\def\mnras{\aaref@jnl{Mon.~Not.~Roy.~Astron.~Soc.}}             
\def\prd{\aaref@jnl{Phys.~Rev.~D}}        
\def\prc{\aaref@jnl{Phys.~Rev.~C}}  
\def\prl{\aaref@jnl{Phys.~Rev.~Lett.}}    
\def\qjras{\aaref@jnl{QJRAS}}             
\def\skytel{\aaref@jnl{S\&T}}             
\def\ssr{\aaref@jnl{Space~Sci.~Rev.}}     
\def\zap{\aaref@jnl{ZAp}}                 
\def\nat{\aaref@jnl{Nature}}              
\def\aplett{\aaref@jnl{Astrophys.~Lett.}} 
\def\apspr{\aaref@jnl{Astrophys.~Space~Phys.~Res.}} 
\def\physrep{\aaref@jnl{Phys.~Rep.}}      
\def\physscr{\aaref@jnl{Phys.~Scr}}       
\def\commat{\aaref@jnl{Comm.~Math.~Phys.}}              
\def\science{\aaref@jnl{Science}}               
\def\cqg{\aaref@jnl{Classical Quant.~Grav.}}            
\def\jpcs{\aaref@jnl{JPCS}}                                     
\def\ijmpd{\aaref@jnl{Int.~J.~Mod.~Phys.~D}}                    
\def\grg{\aaref@jnl{Gen.~Relat.~Gravit.}}               
\def\rpp{\aaref@jnl{Rep.~Prog.~Phys.}}          
\def\npa{\aaref@jnl{Nucl.~Phys.~A}}        
\def\lrr{\aaref@jnl{Living Rev.~Rel.}}                   
\def\jcap{\aaref@jnl{J.~Cosmology Astropart.~Phys.}}    
\def\rmp{\aaref@jnl{Rev.~Mod.~Phys.}}   
\def\epjc{\aaref@jnl{Eur.~Phys.~J.~C}} 
\def\plb{\aaref@jnl{~Phy.~Lett.~B}} 
\def\mpla{\aaref@jnl{Mod.~Phy.~Lett.~A}} 
\def\arxiv{\aaref@jnl{arxiv.org}}

\allowdisplaybreaks[1]

\begin{document}
\title{Cosmological Scenarios in $f(Q,C)$ gravity with a dynamical degree of freedom}
\author{Hamid Shabani\orcidlink{0000-0002-2309-3591}}
\email{h.shabani@phys.usb.ac.ir}
\affiliation{Physics Department, Faculty of Sciences, University of Sistan and Baluchestan, Zahedan, Iran}
\author{Avik De\orcidlink{0000-0001-6475-3085}}
\email{avikde@um.edu.my}
\affiliation{Institute of Mathematical Sciences, Faculty of Science, Universiti Malaya, 50603 Kuala Lumpur, Malaysia}



\begin{abstract}
The $f(Q,C)$ theory, which extends symmetric teleparallel gravity by including the boundary term $C$ in addition to the non-metricity scalar $Q$, provides a unifying framework that encompasses both $f(Q)$ and $f(\mathring{R})$ gravities. In this work, we develop a comprehensive dynamical system analysis of $f(Q,C)$ cosmology formulated within the non-coincident affine connection branches. Unlike the coincident case, these branches introduce a dynamical degree of freedom that significantly enriches the cosmological phase space. We show that even for simple power-law forms of the Lagrangian, the theory accommodates a broad spectrum of cosmic scenarios, including successive pressureless matter eras, stiff-matter phases between early inflation and dark matter domination, and late-time acceleration. Our analysis demonstrates that the Universe can naturally evolve from an initial de Sitter phase to a matter-dominated epoch and subsequently to a final de Sitter attractor, consistent with the observed thermal history. These results highlight the role of the dynamical connection in shaping cosmic evolution and underline the potential of $f(Q,C)$ gravity as a viable alternative framework for addressing outstanding issues in modern cosmology.
\end{abstract}

\maketitle

\section{Introduction}
General Relativity (GR) has been remarkably successful in describing the gravitational interaction across astrophysical and cosmological scales. Nevertheless, theoretical challenges such as non-renormalizability, the cosmological constant problem, and the inability to fully account for cosmic acceleration continue to motivate the search for viable modifications of GR \cite{heisenberg}. A fruitful direction arises within the framework of teleparallel and symmetric teleparallel geometries, where torsion or non-metricity, rather than curvature, mediates the gravitational interaction.

In particular, the symmetric teleparallel equivalent of General Relativity (STGR) is formulated on a curvature-free and torsion-free manifold, with dynamics governed by the non-metricity scalar $Q$. While STGR is dynamically equivalent to GR at the level of equations, modifications such as the scalar-non-metricity gravity \cite{laur,ghulamqphi1,ghulamqphi2} or $f(Q)$ gravity \cite{coincident,fqsurvey,ghulamfqbi} introduce non-trivial extensions with novel cosmological implications. A further generalization can be achieved by incorporating the boundary term $C$, defined via the relation $\mathring{R}=Q+C$, leading to the recently proposed $f(Q,C)$ gravity \cite{als/fQC, Capozziello2023a}. Such a construction embeds both $f(Q)$ and $f(\mathring{R})$ theories as special cases, while offering a richer phenomenology due to the independent treatment of $Q$ and $C$.

At the cosmological level, $f(Q,C)$ gravity yields modified Friedmann equations that effectively introduce a geometrical dark energy sector. Previous studies have demonstrated that this framework can reproduce the observed thermal history of the universe, including transitions between matter and dark-energy dominated epochs, as well as dynamical behaviors such as phantom divide crossing. Several recent works systematically analyzed the cosmological implications, highlighting the role of boundary contributions in shaping cosmic evolution \cite{shabani1:2025, ghulamannals, samaddar2025, Sadatian2024, Sharif2025, Alruwaili2025, ppaliathanasis2024, Muhammad2024, BH, bmishra, recon, ganeshpert}.

A crucial feature of symmetric teleparallel cosmology is the existence of distinct affine connection classes consistent with homogeneity and isotropy of the Friedmann--Lemaître--Robertson--Walker (FLRW) Universe. Beyond the trivial coincident gauge (Connection I)\footnote{It has been demonstrated that this class of connections yields cosmological dynamics equivalent to those already established in the older $f(T,B)$ gravity formulated within the metric teleparallel approach \cite{baha23023}. Consequently, its implications are well understood, and a further detailed analysis here would be redundant}, two further classes of affine connections (Connection II and Connection III) admit an additional dynamical degree of freedom, represented by a function $\gamma(t)$. 
Recent progress highlights the need to probe general non-coincident connection branches, whose extra degrees of freedom can imprint nontrivial signatures on cosmological dynamics. This broader setting enlarges the solution space and may uncover phenomenology that remains hidden in the coincident gauge. A series of exact cosmological solutions in non-coincidence $f(Q)$ gravity has been identified recently \cite{avikand}. A dynamical system treatment of $f(Q)$ cosmology beyond the coincident branch has been developed in \cite{Paliathanasis:2023nkb,shabanifQ,Narawade:2024pxb,ghulamfQ,saikatfQ}, where a consistent succession of cosmic epochs was exhibited. Building on this direction, \cite{rajadata} examined a particular non-coincident realization specified via a Hubble-based parameterization, while \cite{saridakisdata} employed Hubble and Gaussian–process reconstructions to infer the dynamical connection degree of freedom for two widely studied $f(Q)$ models. At the background level, $\Lambda$CDM-like evolutions arising from non-coincident connections were analyzed in \cite{saikatfQlcdm}, enabling analytic reconstruction for connection class~II and a numerical reconstruction for class~III, under a cosmographic prior. On the other hand, the non-coincident formulation of power-law $f(Q)$ gravity was shown to challenge $\Lambda$CDM from DESI DR2 \cite{an01}.

When expressed as a function of the scale factor $a(t)$, this quantity introduces new dynamical effects that can mediate interactions between dark matter and dark energy sectors, thereby influencing the global phase-space structure of the theory.

In this work, we perform a comprehensive dynamical system analysis of $f(Q,C)$ cosmology formulated within Connection classes II and III, incorporating the scale-factor-dependent degree of freedom $\gamma(t)$. Our aim is to investigate the stability of cosmological fixed points, the viability of standard evolutionary sequences (radiation, matter, and dark-energy domination), and the possible alleviation of cosmological tensions through the dynamical interplay between $Q$ and $C$. This study extends earlier works by providing a unified dynamical framework for non-coincident symmetric teleparallel connections, thereby revealing the full range of cosmological behavior admitted by $f(Q,C)$ gravity.
\section{Basic formulation of $f(Q,C)$ gravity}\label{formulation}
In the framework of symmetric teleparallel theories of gravity, the affine connection is taken to be torsion-free and curvature-free, while the gravitational interaction is entirely encoded in the non-metricity of spacetime. The latter is quantified by the failure of the metric tensor to be covariantly conserved, which is expressed through the non-metricity tensor
\begin{equation} \label{Q tensor}
Q_{\lambda\mu\nu} := \nabla_\lambda g_{\mu\nu}=\partial_\lambda g_{\mu\nu}-\Gamma^{\beta}_{\,\,\,\mu\lambda}g_{\beta\nu}-\Gamma^{\beta}_{\,\,\,\nu\lambda}g_{\beta\mu}\neq 0,
\end{equation}
signalling that lengths and angles are not preserved under parallel transport. In this way, non-metricity plays the central role in mediating gravity, as opposed to curvature in General Relativity or torsion in teleparallel $f(T)$ gravity.

A generic affine connection $\Gamma$ can be decomposed into the Levi-Civita connection $\mathring \Gamma$ and the disformation tensor $L^\lambda{}_{\mu\nu}$ as
\begin{equation} \label{connc}
\Gamma^\lambda{}_{\mu\nu} := \mathring{\Gamma}^\lambda{}_{\mu\nu}+L^\lambda{}_{\mu\nu},
\end{equation}
where the disformation encodes the pure effects of non-metricity. Explicitly,
\begin{equation} \label{L}
L^\lambda{}_{\mu\nu} = \frac{1}{2} \left( Q^\lambda{}_{\mu\nu} - Q_\mu{}^\lambda{}_\nu - Q_\nu{}^\lambda{}_\mu \right) \,.
\end{equation}

Two independent non-metricity vectors can be constructed as
\begin{equation*}
 Q_\mu := g^{\nu\lambda}Q_{\mu\nu\lambda} = Q_\mu{}^\nu{}_\nu \,, \qquad 
 \tilde{Q}_\mu := g^{\nu\lambda}Q_{\nu\mu\lambda} = Q_{\nu\mu}{}^\nu \,,
\end{equation*}
with corresponding contractions of the disformation tensor
\begin{align}
 L_\mu := L_\mu{}^\nu{}_\nu \,, \qquad 
 \tilde{L}_\mu := L_{\nu\mu}{}^\nu \,.   
\end{align}

The superpotential (or non-metricity conjugate) tensor $P^\lambda{}_{\mu\nu}$ is defined as
\begin{equation} \label{P}
P^\lambda{}_{\mu\nu} = \frac{1}{4} \left( -2 L^\lambda{}_{\mu\nu} + Q^\lambda g_{\mu\nu} - \tilde{Q}^\lambda g_{\mu\nu} -\delta^\lambda{}_{(\mu} Q_{\nu)} \right) \,,
\end{equation}
which allows one to construct the non-metricity scalar
\begin{equation} \label{Q}
Q=Q_{\alpha\beta\gamma}P^{\alpha\beta\gamma}\,.
\end{equation}

The above geometrical quantities satisfy the following identities:
\begin{align}
\m R_{\mu\nu}+\m\nabla_\alpha L^\alpha{}_{\mu\nu}-\m\nabla_\nu\tilde L_\mu
+\tilde L_\alpha L^\alpha{}_{\mu\nu}-L_{\alpha\beta\nu}L^{\beta\alpha}{}_\mu=0\,,
\label{mRicci}\\
\m R+\m\nabla_\alpha(L^\alpha-\tilde L^\alpha)-Q=0\,. \label{mR}
\end{align}
Since $Q^\alpha-\tilde Q^\alpha=L^\alpha-\tilde L^\alpha$, one can define the boundary term
\begin{align}\label{aaboundary}
C=\m{R}-Q&=-\m\nabla_\alpha(Q^\alpha-\tilde Q^\alpha) \notag \\
&=-\frac1{\sqrt{-g}}\partial_\alpha\left[\sqrt{-g}(Q^\alpha-\tilde Q^\alpha)\right].
\end{align}
This boundary term provides the link between the Ricci scalar $\m R$ of General Relativity and the non-metricity scalar $Q$.

Motivated by these relations, the $f(Q,C)$ theory extends both $f(Q)$ and $f(\m R)$ gravities into a single unified framework. Its action is given by
\begin{equation}
S=\int \left[ \frac{1}{2\kappa }f(Q,C)+\mathcal{L}_{M}\right] \sqrt{-g}\,d^{4}x\,,
\label{eqn:action-fQC}
\end{equation}
where $f$ is an arbitrary function of $Q$ and $C$, and $\mathcal L_M$ denotes the matter Lagrangian. Variation with respect to the metric yields the field equations
\begin{align}\label{eqn:FE1}
\kappa T_{\mu\nu}
=&-\frac f2g_{\mu\nu}+2P^\lambda{}_{\mu\nu}\nabla_\lambda(f_Q-f_C)
  +\left(\m G_{\mu\nu}+\frac Q2g_{\mu\nu}\right)f_Q \notag\\
  &+\left(\frac C2g_{\mu\nu}-\m\nabla_{\mu}\m\nabla_{\nu}
  +g_{\mu\nu}\m\nabla^\alpha\m\nabla_\alpha \right)f_C\,.
\end{align}
Since the affine connection is independent of the metric tensor in this framework, variation of the action with respect to the connection leads to the second set of field equations,
\begin{align}\label{eqn:FE2-invar}
(\nabla_\mu-\tilde L_\mu)(\nabla_\nu-\tilde L_\nu)
\left[4(f_Q-f_C)P^{\mu\nu}{}_\lambda\right]=0\,,
\end{align}
in the absence of hypermomentum.

In what follows, we aim to investigate the cosmological implications of $f(Q,C)$ gravity. For this purpose, we adopt the spatially flat Friedmann–Robertson–Walker (FLRW) geometry,
\begin{equation}
ds^2=-dt^2+a^2(t)\left(dr^2+r^2d\theta^2+r^2\sin^2\theta d\phi^2\right),  \label{3a}
\end{equation}
corresponding to a homogeneous and isotropic matter distribution. Here $a(t)$ is the cosmic scale factor, and its time derivative defines the Hubble parameter $H(t)=\dot{a}(t)/a(t)$, with $\dot{()}$ denoting differentiation with respect to cosmic time $t$.

The FLRW metric (\ref{3a}) exhibits cosmic symmetry, specifically homogeneity and isotropy, which can be described by spatial rotational and translational transformations. A symmetric teleparallel affine connection $\Gamma$ is an affine connection that is both torsion-free and curvature-free. It also possesses both spherical and translational symmetries, implying vanishing of the Lie derivative of the affine connection in terms of the Killing vectors. There exist three classes of such affine connections characterized by a dynamic degree of freedom $\gamma(t)$, as outlined in the references \cite{FLRW/connection, ad/bianchi}
\begin{align} 
\Gamma^t{}_{tt}=&C_1, 
	\quad 					\Gamma^t{}_{rr}=C_2, 
	\quad 					\Gamma^t{}_{\theta\theta}=C_2r^2, 
	\quad						\Gamma^t{}_{\phi\phi}=C_2r^2\sin^2\theta,								\notag\\
\Gamma^r{}_{tr}=&C_3, 
	\quad  	\Gamma^r{}_{rr}=0, 
	\quad		\Gamma^r{}_{\theta\theta}=-r, 
	\quad		\Gamma^r{}_{\phi\phi}=-r\sin^2\theta,												\notag\\
\Gamma^\theta{}_{t\theta}=&C_3, 
	\quad		\Gamma^\theta{}_{r\theta}=\frac1r,
	\quad		\Gamma^\theta{}_{\phi\phi}=-\cos\theta\sin\theta,										\notag\\
\Gamma^\phi{}_{t\phi}=&C_3, 
	\quad 	\Gamma^\phi{}_{r\phi}=\frac1r, 
	\quad 	\Gamma^\phi{}_{\theta\phi}=\cot\theta.
\end{align}
\begin{enumerate}\label{casesConn}
\item[(I)]  $C_1=\gamma$, $C_2=C_3=0$; or
\item[(II)] $C_1=\gamma+\dfrac{\dot\gamma}\gamma$, $C_2=0$, $C_3=\gamma$,
             where $\gamma$ is non-vanishing; or 
\item[(III)] $C_1=-\dfrac{\dot\gamma}{\gamma}$, $C_2=\gamma$, $C_3=0$,
             where $\gamma$ is non-vanishing.
\end{enumerate} 
Let us denote these classes by Connection type $\Gamma_I$, $\Gamma_{II}$ and $\Gamma_{III}$, respectively. For specific cosmological applications, it is necessary to analyze each class of these symmetric teleparallel connections and establish the associated equations of motion for $f(Q,C)$ gravity, which were derived in \cite{als/fQC}. We analyse each phenomenon in the following sections for the functional form $f(Q,C)=f_{1}(Q)+f_{2}(C)$  and $\gamma=s(a)$ . As described in the footnote above, we refrain from discussing the connection class $\Gamma_I$.

\section{Connection type $II$}\label{type2}
The equations of motion for this class of affine connections take the form~\cite{als/fQC}
 
\begin{align}
&\kappa \rho_m=\frac{f}{2}+f_Q \left(3 H^2-\frac{Q}{2}\right)-\frac{C f_C}{2}+\frac{3 \gamma  \dot{f}_Q}{2}+\left(3 H-\frac{3 \gamma}{2}\right)\dot{f}_C,\label{t2.1}\\
&\kappa p_m=-\frac{f}{2}+f_Q \left(-3 H^2-2 H'+\frac{Q}{2}\right)+\frac{C f_C}{2}+\frac{1}{2} (3 \gamma -4 H) \dot{f}_Q-\frac{3 \gamma  \dot{f}_C}{2}-\ddot{f}_C,\label{t2.2}\\
&\kappa\m\nabla_{\mu}T^{\mu}_{t}=\mathcal{C}_{II},\label{t2.2.1}\\
&\mathcal{C}_{II}=\frac{3}{2}\gamma\Big[3H\dot{f}_Q+\ddot{f}_Q-3H\dot{f}_C-\ddot{f}_C\Big],\nonumber
\end{align}

where we define

\begin{align}
&Q=-6 H^2+9 \gamma  H+3 \dot{\gamma},\label{t2.3}\\
&C=6 \left(3 H^2+\dot{H}\right)-9 \gamma  H-3 \dot{\gamma},\label{t2.4}\\
&Q+C=6(2H^{2}+\dot{H})\label{t2.4.1}.
\end{align}

In the case of $f(Q,C)=f_{1}(Q)+f_{2}(C)$  and $\gamma=s(a)$ we introduce the following dimensionless variables

\begin{align}\label{t2.5}
&\Omega_{m}=\frac{\kappa\rho_{m}}{3H^2f_{1}'},~~~x_1=\frac{Q}{H^2},~~~x_2=\frac{C}{H^2},~~~x_3=\frac{\dot{H}}{H^2},~~~x_4=\frac{\gamma }{H},\nonumber\\
&x_5=\frac{2 H\dot{Q}}{Q^2},~~~x_6=\frac{2 H\dot{C}}{C^2},~~~x_7=\frac{f_{2}}{f_{1}},~~~x_8=\frac{f_{1}}{x_{1}}
\end{align}

and also introduce the following parameterizations

\begin{align}\label{t2.6}
&j=\frac{Q f_{1}'}{f_{1}},~~~d=\frac{Q f_{1}''}{f_{1}'},\\\nonumber
&i=\frac{C f_{2}'}{ f_{2}},~~~e=\frac{C  f_{2}''}{ f_{2}'},~~~p=\frac{C  f_{2}'''}{ f_{2}''},\\\nonumber
&n=\frac{a s'}{s},~~~m=\frac{a s''}{s'},~~~l=\frac{a s'''}{s''},
\end{align}

where primes denote differentiation with respect to the argument. We now proceed to derive the dynamical system equations of motion as follows; first, the following results are obtained from eq.~(\ref{t2.1}) and eq.~(\ref{t2.2}), respectively

\begin{align}
&\Omega_{{\sf m}}=1-\Omega_{{\sf DE}} =\frac{1}{12 j}\bigg\{j \Big[x_1 \left(3 d x_4 x_5-2\right)+12\Big]+x_1 \Big[x_7 \Big(i \big(-3 e \left(x_4-2\big) x_6-2\right)+2\Big)+2\Big]\bigg\}\label{t2.7}\\
&4\frac{\ddot{C}}{H^{4}}=\frac{x_2^2 }{e i x_1 x_7}\Bigg[j \bigg(d x_1 x_5 \big(-3 (w-1) x_4-4\big)+2 (w+1) \left(x_1-6\right)-8 x_3\bigg)-2 (w+1) x_1+\nonumber\\
& x_1 x_7\bigg(-e i p x_2 x_6^2+3 e i x_6 \big(w \left(x_4-2\big)-x_4\right)+2 (i-1) (w+1)\bigg)\Bigg],\label{t2.8}
\end{align}

where the subscript ``${\sf m}$" denotes the contribution to the matter content of the Universe, while ``${\sf DE}$" represents the contribution from dark energy. It should be noted that eq.~(\ref{t2.8}) is absent when $e=0$. Next, we derive the required orders of time derivatives for the variables $Q$, $H$ and $\gamma$. Equating $\dot{\gamma}$ extracted from eq.~(\ref{t2.3}) and eq.~(\ref{t2.4}) yields the following constraint

\begin{align}\label{t2.9}
x_{1}+x_{2}=6 (x_{3}+2).
\end{align}

The second time derivative of $\gamma$ is calculated using either eq.~(\ref{t2.3}) or eq.~(\ref{t2.4}) , which, when considering~eq.~(\ref{t2.9}) leads to the same relation, and is given by

\begin{align}\label{t2.10}
\ddot{\gamma}=\frac{1}{6} \Big[x_1^2 x_5-6 \big(x_1-4 x_3+3 \left(x_3-3\right) x_4+6\big)\Big]H^3 .
\end{align}

Taking the time derivative of both sides of~eq.~(\ref{t2.4.1}) gives

\begin{align}\label{t2.11}
\ddot{H}=-\frac{1}{12} H^3 \left(-x_5 x_1^2+48 x_3-x_2^2 x_6\right).
\end{align}

For the third time derivative of $\gamma$ we again use either eq.~(\ref{t2.3}) or eq.~(\ref{t2.4}). Equation~(\ref{t2.3}) then gives

\begin{align}\label{t2.12}
&\dddot{\gamma}=\frac{1}{12} \Bigg\{H^4 \bigg[+\left(36-24 x_3\right) x_1+12 \Big(-27 x_4+x_3 \left(4 x_3+39 x_4-40\right)+18\Big)+\\\nonumber
&-\Big(\left(3 x_4+2\right) x_5 x_1^2\Big)+ \left(4-3 x_4\right) x_6x_2^2\bigg]+ 4 \ddot{Q}\Bigg\}
\end{align}

which includes $\ddot{Q}$. Also, eq.~(\ref{t2.4}) leads to

\begin{align}\label{t2.13}
&\dddot{\gamma}=\frac{H^{4}}{12 e i x_1 x_7}\Bigg\{x_2^2 \bigg[j \Big(d x_1 x_5 \big(3 (w-1) x_4+4\big)-2 (w+1) \left(x_1-6\right)+8 x_3\Big)+2 (w+1) x_1\bigg]+\\\nonumber
&x_1 x_7 \bigg[e i \Big(p x_6^2 x_2^3+3 x_6 x_2^2 \big(4+w(2-x_4)\big)+12 \left(2 x_3-3\right) x_2+3 \left(x_4-2\right) \left(-x_5 x_1^2+156 x_3-108\right)\Big)-\\\nonumber
&2 (i-1) (w+1) x_2^2\bigg]\Bigg\}+2\dddot{H}.
\end{align}
which this form of $\dddot{\gamma}$ contains $\dddot{H}$. Hence, equating eq.~(\ref{t2.12}) with eq.~(\ref{t2.13}) provides a relation between $\ddot{Q}$ and $\dddot{H}$. Moreover, the particular case $\gamma=s(a)$ yields additional relations, which are

\begin{align}
&\dot{\gamma}=n x_4 H^2\label{t2.14}\\
&\ddot{\gamma}=H^3 n x_4 \left(m+x_3+1\right), \label{t2.15}\\
&\dddot{\gamma}=\frac{H^4 n x_4}{12}  \Big[12 m \big(l+3 x_3+3\big)+x_5 x_1^2-12 x_3+x_2^2 x_6+12\Big]. \label{t2.16}
\end{align}

Combining expressions from eq.~(\ref{t2.14})-(\ref{t2.16}) for the time derivatives of $\gamma$ with those derived above yields two additional constraints

\begin{align}
&3 (n+3) x_4+x_2=6 \left(x_3+3\right),\label{t2.17}\\
&2 n x_4 \left(m+x_3+4\right)+6 x_3 x_4=3 x_5 \big[(n+3) x_4-2\big]^2+8 x_3\label{t2.18}.
\end{align}

Hence, $\ddot{Q}$ is obtained by identifying $\dddot{\gamma}$ as it appears in eqs.~(\ref{t2.12}) and~(\ref{t2.16}). This gives

\begin{align}\label{t2.19}
&4\frac{\ddot{Q}}{H^4 }=12 x_4 \bigg[m n \left(l+3 x_3+3\right)-(n+39)x_3+n+27\bigg]+12 x_1 \left(2 x_3-3\right)-24 \bigg[2 \left(x_3-10\right) x_3+9\bigg]+\nonumber\\
&x_5 x_1^2 \big((n+3) x_4+2\big)+x_2^2 x_6 \big((n+3) x_4-4\big)
\end{align}

and a similar procedure is applied to eqs.~(\ref{t2.13}) and~(\ref{t2.16}), which results in

\begin{align}\label{t2.20}
&\dddot{H}=\frac{H^4}{24 e i x_1 x_7}\Bigg\{x_2^2 \bigg[d j x_1 x_5 \Big(-3 (w-1) x_4-4\Big)+2 (w+1) \Big(j \left(x_1-6\right)-x_1\Big)-8 j x_3\bigg]+\nonumber\\
&x_1 x_7 \bigg[e i \Big(12 x_4 \big(m n \left(l+3 x_3+3\right)+n \left(-x_3\right)-39 x_3+27\big)+72 \left(13 x_3-9\right)+x_2 \left(36-24 x_3\right)+\nonumber\\
&x_6 x_2^2 \big(x_4 (n+3 w)-6 (w+2)\big)+x_1^2 x_5 \big((n+3) x_4-6\big)-p x_6^2 x_2^3\Big)\bigg]\Bigg\}.
\end{align}

We are now prepared to obtain the equivalent dynamical system. The system corresponding to eqs.(\ref{t2.1})-(\ref{t2.2}) is given by the following relations

\begin{align}
&\frac{dx_{1}}{dN}=\frac{1}{2} x_1 \big(x_1 x_5-4 x_3\big),\label{t2.21}\\
&\frac{dx_{2}}{dN}=\frac{1}{2} x_2 \big(x_2 x_6-4 x_3\big),\label{t2.22}\\
&\frac{dx_{3}}{dN}=\frac{1}{12} \Big(x_5 x_1^2-24 x_3 \left(x_3+2\right)+x_2^2 x_6\Big),\label{t2.23}\\
&\frac{dx_{4}}{dN}=\frac{1}{3} \Big(x_1-3 \left(x_3+3\right) x_4+6\Big),\label{t2.24}\\
&\frac{dx_{5}}{dN}=\frac{1}{2 x_1^2}\Bigg\{12 x_4 \Big[m n \left(l+3 x_3+3\right)+n \left(1-x_3\right)-39 x_3+27\Big]+12 x_1 \left(2 x_3-3\right)-24 \Big[2 \left(x_3-10\right) x_3+9\Big]+\nonumber\\
&x_5 x_1^2 \Big[(n+3) x_4+2 x_3+2\Big]+x_2^2 x_6 \Big[(n+3) x_4-4\Big]-2 x_5^2 x_1^3\Bigg\},\label{t2.25}\\
&\frac{dx_{6}}{dN}=\frac{1}{2 e i x_1 x_7}\Bigg\{j \bigg[d x_1 x_5 \Big(-3 (w-1) x_4-4\Big)+2 (w+1) \left(x_1-6\right)-8 x_3\bigg]-2 (w+1) x_1+\nonumber\\
&x_1 x_7 \bigg[-e i (p+2) x_2 x_6^2+e i x_6 \Big(3 w \left(x_4-2\right)+2 x_3-3 x_4\Big)+2 (i-1) (w+1)\bigg]\Bigg\},\label{t2.26}\\
&\frac{dx_{7}}{dN}=\frac{1}{2} \Big(i x_2 x_6 x_7-j x_1 x_5 x_7\Big),\label{t2.27}\\
&\frac{dx_{8}}{dN}=\frac{x_8 \Bigg\{x_5 \bigg[j x_1^2-\Big(x_2-6 \left(x_3+2\right)\Big){}^2\bigg]-4 x_2 x_3+24 \left(x_3+2\right) x_3\Bigg\}}{2 x_1}.\label{t2.28}
\end{align}

where we have used relation~(\ref{t2.11}) to obtain $dx_{3}/dN$, result~(\ref{t2.19}) to obtain $dx_{5}/dN$ and equation~(\ref{t2.8}) for $dx_{6}/dN$. Equations (\ref{t2.21})-(\ref{t2.28}) constitute the most general form of the cosmological equations of motion in $f(Q,C)=f_{1}(Q)+f_{2}(C)$ gravity, under study of the specific case $\gamma=s(a)$. Under these conditions, constraints~(\ref{t2.9}),~(\ref{t2.17}) and~(\ref{t2.18}) reduce the number of independent variables from eight to five. This reduction enables the elimination of three variables among $x_{1}-x_{5}$. Accordingly, we focus on the phase-space evolution of the Universe in terms of $x_{3}$, $x_{5}$ and $x_{6}$-$x_{8}$ while $x_{1}$, $x_{2}$ and $x_{4}$ are omitted. It should be noted that relation~(\ref{t2.9}) holds regardless of the choice of $\gamma$, whereas equations~(\ref{t2.17})-(\ref{t2.18}) are valid only for $\gamma=s(a)$. \\ 

One might be tempted to analyze~(\ref{t2.21})-(\ref{t2.28}) in their current form; however, this approach is not advisable. As the parameters $j,d,i,e,p,n,m,l$ can be functions of the $x_{i}$ variables, this dependency may influence the results of the analysis\footnote{A similar, less desirable approach was adopted in~\cite{Amendola:2007}.}. Furthermore, the above system may not be suitable for arbitrary~$f(Q,C)$ functions, as the variables defined in~(\ref{t2.5}) and~(\ref{t2.6}) may be insufficient or inconsistent in both number and definition.

With the necessary tools in place, we now investigate particular choices of $f_{1}(Q)$ and $f_{2}(C)$, focusing on power-law forms for both functions. When these functions share the same power, the phase-space dimension is four, and hence, the discussion is thus organized into two categories. In addition, we adopt $\gamma=\gamma_{0} a^{r}$ abd $w=0$ throughout the paper.


\subsection{$f(Q,C)=\alpha Q^{u}+\beta C^{v}$ with $u\neq v$}\label{type2.1}

Here, the system that characterizes the Universe's cosmological behavior is derived by substituting

\begin{align}\label{t2.29}
j=u,~~~d=u - 1,~~~i=v,~~~e=v-1,~~~p=v-2,~~~n=r,~~~m=r-1,~~~l=r-2,
\end{align}

into eqs.~(\ref{t2.21})-(\ref{t2.28}).  Also, using the above results one gets
\begin{align}\label{t2.29.1}
\mathcal{C}_{II}=\frac{3 }{8}H  x_1 x_4 x_8\Bigg\{u(u-1)\bigg[\frac{4 \ddot{Q}}{H^4 x_1^2}+x_5 \Big((u-2) x_1 x_5+6\Big)\bigg]+v(1-v) x_7 \bigg[\frac{4 \ddot{C}}{H^4 x_2^2}+x_6 \Big((v-2) x_2 x_6+6\Big)\bigg]\Bigg\},
\end{align}
for the right hand side of the conservation equation~(\ref{t2.2.1}). In the result~(\ref{t2.29.1}) $\ddot{C}$ and $\ddot{Q}$ are substituted from eqs.~(\ref{t2.3}) and~(\ref{t2.19}). As can be seen, the system cannot be solved straightforwardly; therefore, we first outline the criteria for the existence of different cosmological solutions.We analyze the stability of fixed-point solutions and the corresponding cosmological scenarios in the $(x_{3}, x_{5}, x_{6}, x_{7}, x_{8})$ phase space, since the value of $x_{3}$ determines the type of cosmic era ($w_{{\sf eff}}=-1-2x_{3}/3$) and $dx_{5}/dN$-$dx_{8}/dN$ are sensitive to the model parameters. Furthermore, we adopt the term “true solution/true fixed point” for a critical point that satisfies two criteria: it must have $0\leq\Omega_{{\sf m}}\leq1$ and it must be capable of connecting to other significant fixed points in the phase space, allowing the Universe to transition between different phases.

\subsubsection{Pressure-less matter ({\sf PLM}) solution with $w_{{\sf eff}}=0$}\label{type2.1.1}

A cosmic era dominated by a {\sf PLM} is described by $w_{{\sf eff}}=0$ (equivalently $x_{3}=-3/2$). The set of eqs.~(\ref{t2.21})-(\ref{t2.28}) permits a {\sf PLM} solution with the following characteristics

\begin{align}
&P^{{\sf PLM}}:\left(x_1= -6, x_2= 9, x_3= -\frac{3}{2}, x_4=0, x_5= 1, x_6= -\frac{2}{3}, x_7=0, x_8= 0\right)\label{t2.30}\\ 
&\mbox{Ex.}\to \left(u=1, v=-\frac{1}{2}\right),~\mbox{Ei.}=\left[-3, \frac{9}{2},\frac{27}{2},\frac{1}{4} \left(-\sqrt{\frac{96}{\text{x7}}+9}-9\right),\frac{1}{4} \left(\sqrt{\frac{96}{\text{x7}}+9}-9\right)\right],~\Omega_{{\sf m}}|_{P^{{\sf PLM}}}^{(u=1, v=-\frac{1}{2})}=1,\label{t2.31}\\
& \mbox{Ex.}\to (u=1, v\approx1),~\mbox{Ei.}=\left[-3, 3(1-v), \frac{27}{2}, 3 \sqrt{\frac{1}{2 (v-1) \text{x7}}}+\frac{9}{4},-3 \sqrt{\frac{1}{2 (v-1) \text{x7}}}+\frac{9}{4}\right],~\Omega_{{\sf m}}|_{P^{{\sf PLM}}}^{(u=1, v\approx1)}=1,\label{t2.32}\\
&\mbox{Ex.}\to\left (u=\frac{1}{2}, v=-\frac{1}{2}\right),~\mbox{Ei.}=\left[-3,-\frac{3}{2},-\frac{3}{2},3,\frac{27}{2}\right],~\Omega_{{\sf m}}|_{P^{{\sf PLM}}}^{(u=\frac{1}{2}, v=-\frac{1}{2})}=0,\label{t2.33}\\
& \mbox{Ex.}\to (u=\frac{1}{2}, v\approx1),~\mbox{Ei.}=\left[-3, \frac{3}{2}-3 v, 0, -\frac{1}{4}\left (12 v-5\right) \left(\sqrt{\frac{72}{(v-1) v \text{x7}}+1}-1\right)\right.,\nonumber\\
&\left. -\frac{1}{4}\left (12 v-5\right) \left(-\sqrt{\frac{72}{(v-1) v \text{x7}}+1}-1\right)\right],~\Omega_{{\sf m}}|_{P^{{\sf PLM}}}^{(u=\frac{1}{2}, v=1)}=0\label{t2.34}.
\end{align}


In this context, ``Ex." and ``‘Ei" refer to the existence conditions and eigenvalues, respectively, while $\Omega_{{\sf m}}$is given by eq.~(\ref{t2.5}). The first two solutions are true solutions and always manifest as saddle points; the remaining two do not correspond to a {\sf PLM}–dominated era. Notably, matter domination with $w_{{\sf eff}}=0$ and $\Omega_{{\sf PLM}}=1$ is realized only when $u=1$. These models recover {\sf GR} while also incorporating terms that can justify accelerated expansion in the early and late Universe.


\subsubsection{de Sitter solution with $w_{{\sf eff}}=-1$}\label{type2.1.2}

There is an exact de Sitter solution with $w_{{\sf eff}}=-1$ (or equivalently $x_{3}=0$) under the following conditions

\begin{align}
&P^{{\sf dS}}:\left(x_1= -6, x_2= 18, x_3= 0, x4=0, x_5= 0, x_6=0, x_7=\frac{1-2 u}{v-1}, x_8= 0\right),\label{t2.35}\\
&\mbox{Ex.}\to (u, v),~\mbox{Ei.}=\left[0,-3,3,-\frac{3}{2} \left(\sqrt{9-8 u}+1\right),\frac{3}{2} \left(\sqrt{9-8 u}-1\right)\right],~\Omega_{{\sf DE}}|_{P^{{\sf dS}}}=1.\label{t2.36}
\end{align}

The de Sitter solution exists for arbitrary values of $u$ and $v$ and is an unstable solution.


\subsubsection{Solutions with $w_{{\sf eff}}=-1-\frac{2}{3}r$}\label{type2.1.3}

By expressing eqs.~(\ref{t2.21})-(\ref{t2.28}) in terms of the five-dimensional phase-space variables, one finds that $dx_{1}/dN$ and $dx_{4}/dN$ are proportional to $(r-x_3)$. This observation motivates checking whether $x_3=r$ constitutes a general solution.Indeed, such a solution exists, corresponding to $w_{{\sf eff}}=-1-2r/3$, with the following properties 

\begin{align}
&P^{{\sf X}}:\left(x_1= 12+6r, x_2= 0, x_3= r, x_4=2, x_5=\frac{2 r}{3 (r+2)}, x_6= -\frac{1}{(r-3) v}, x_7=0, x_8= 0\right)~~\mbox{provided}~~r=\frac{3 (2-u)}{2 u^2-u-3},\label{t2.37}\\
&\mbox{Ex.}\to (u, v),~\mbox{Ei.}=\left[-\frac{6 (u-2) u}{(u+1) (2 u-3)},\frac{6 (u-2) u}{(u+1) (2 u-3)},\frac{15-6 u^2}{2 u^2-u-3},\frac{3 \left(\sqrt{u \left(4 u^3-15 u-20\right)+45}-5 u+10\right)}{-2 u^2+u+3},\right.\nonumber\\
&\left.\frac{3 \left(\sqrt{u \left(4 u^3-15 u-20\right)+45}+5 u-10\right)}{(u+1) (2 u-3)}\right],\label{t2.38}\\
&~\Omega_{{\sf m}}|_{P^{{\sf X}}}^{r=\frac{3 (2-u)}{2 u^2-u-3}}=\frac{2 (u-2) (4 u-5)}{-2 u^2+u+3}.\label{t2.39}
\end{align}

Depending on the values of $r$, $u$ and $v$, this solution can describe a {\sf PLM}, dark energy, inflationary, {\sf URM}, or {\sf SM} epoch, as we illustrate in the subsequent sections.


\subsubsection{Dark Energy solutions with $w_{{\sf eff}}<-1/3$}\label{type2.1.4}

For appropriate intervals of $r$ the condition $w_{{\sf eff}}<-1/3$ is satisfied, so that $P^{{\sf X}}$ effectively mimics a dark energy solution. Consequently, there exists a solution corresponding to an accelerated expansion epoch, with properties listed below

\begin{align}
P^{{\sf DE}}=P^{{\sf x}}~~~\mbox{provided}~~~ r>-1 ~~~\mbox{or equivalently}~~~ u<-1\lor u>\frac{3}{2}.\label{t2.44}
\end{align}

It is clear that $P^{{\sf DE}}$ is unstable due to the first two eigenvalues. Specifying $u$ allows $r$ to be determined via~(\ref{t2.37}); specifically, $w_{{\sf eff}}<-1/3$ implies $u<-1$ or $u>\frac{3}{2}$. Thus, $P^{{\sf DE}}$ cannot represent an accelerated expansion era (early or late) connected to the {\sf PLM} solution ($u=1$ or $u=1/2$). As a result, $P^{{\sf DE}}$ is not a true solution, and a cosmic evolution sequence ${\sf DE}\to {\sf PLM} \to {\sf dS}$ incorporating both early inflation and late-time acceleration is not realized.


\subsubsection{{\sf URM} solutions with $w_{{\sf eff}}=1/3$}\label{type2.1.5}

Our study indicates that fixed points representing the {\sf URM} epoch exist under the following conditions

\begin{align}
&P^{{\sf Rad}}:\left(x_1= -6, x_2= 6, x_3= -2, x_4=0, x_5=\frac{4}{3},x_6= -\frac{4}{3}, x_7=0, x_8= 0\right),\label{t2.43}\\
&\mbox{Ex.}\to (u=\frac{1}{2}, v=-\frac{1}{4}),~\mbox{Ei.}=\left[-2,17,13, \mathfrak{a}_{1}, \mathfrak{a}_{2}\right],~\Omega_{{\sf m}}|_{P^{{\sf Rad}}}^{(u=\frac{1}{2}, v=-\frac{1}{4})}=0.\label{t2.44}\\
&\mbox{Ex.}\to (u=\frac{1}{2}, v=\frac{3}{4}),~\mbox{Ei.}=\left[-2,17,29, \mathfrak{a}_{3}, \mathfrak{a}_{4}\right],~\Omega_{{\sf m}}|_{P^{{\sf Rad}}}^{(u=\frac{1}{2}, v=\frac{3}{4})}=0,\label{t2.45}\\
&\mbox{Ex.}\to (u=\frac{3}{4}, v=-\frac{1}{4}),~\mbox{Ei.}=\left[-3,17,4, \mathfrak{a}_{5}, \mathfrak{a}_{6}\right],~\Omega_{{\sf m}}|_{P^{{\sf Rad}}}^{(u=\frac{3}{4}, v=-\frac{1}{4})}=\frac{2}{3},\label{t2.46}
\end{align}

where $\mathfrak{a}_{i}$ become as $x_{7}\to 0$. Consequently, these correspond to saddle points characteristic of a radiation-dominated epoch. The first two points are not true solutions due to their matter density parameters, and the third is also not a true solution, as it cannot connect to $P^{{\sf PLM}}$.


\subsubsection{{\sf SM} solutions with $w_{{\sf eff}}=1$}\label{type2.1.6}

In this case we have $w_{{\sf eff}}=1$ (or equivalently $x_{3}=-2$). The system (\ref{t2.29})-(\ref{t2.33}) admits the {\sf SM} solution with

\begin{align}
&P^{{\sf SM}}:\left(x_1= -6, x_2=0, x_3= -3, x_4=0, x_5=-2,x_6= \frac{1}{3 v}, x_7=0, x_8= 0\right),\label{t2.47}\\
&\mbox{Ex.}\to \left(u=\frac{1}{2}, v\right),~\mbox{Ei.}=\left[-3, -3, 3, 24, 6\right],~\Omega_{{\sf m}}|_{P^{{\sf SM}}}^{u=\frac{1}{2}}=0.\label{t2.48}
\end{align}

Consequently, $P^{{\sf SM}}$ does not qualify as a true fixed point, since its matter density parameter vanishes. A summary of the consistent fixed points corresponding to connection type $II$ with $u\neq v$ (in the 5-D phase space) is given in Table~\ref{tab1}. We see that $\mathcal{C}_{II}$ which obtained in~(\ref{t2.29.1}) becomes zero at the fixed-point solutions, which means the system reaches stability when energy is conserved.

\begin{figure}[h!]
\centering
\includegraphics[width=.45\textwidth]{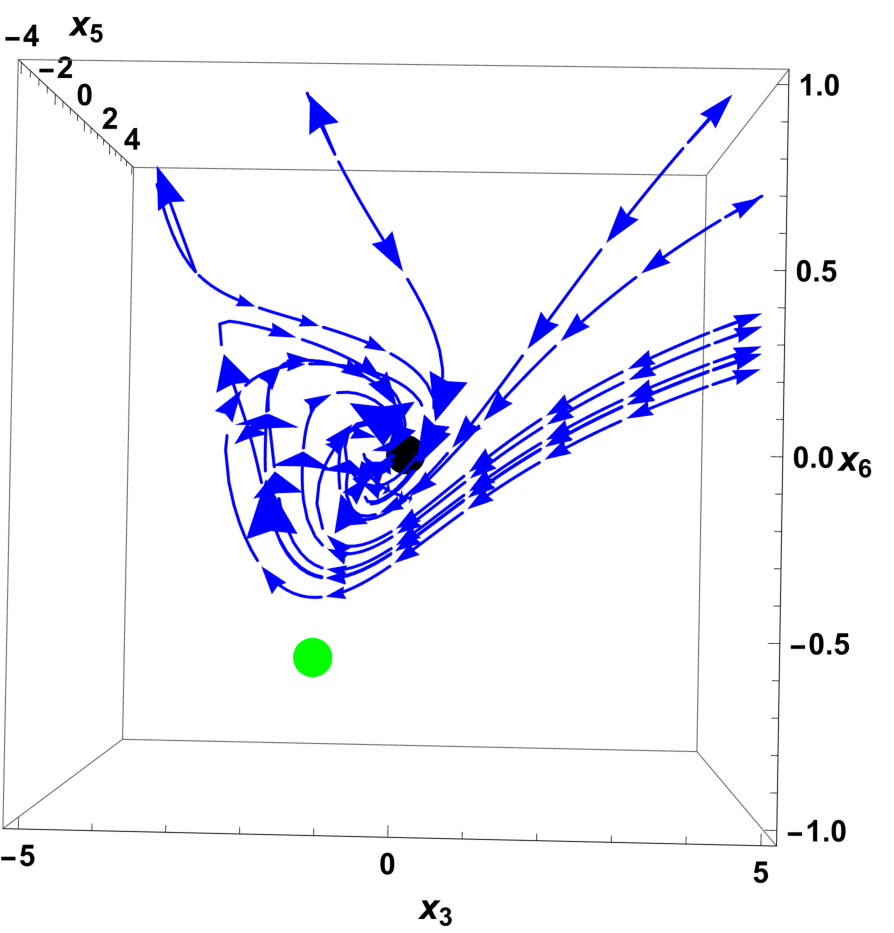}\hspace{2mm}
\includegraphics[width=.4\textwidth]{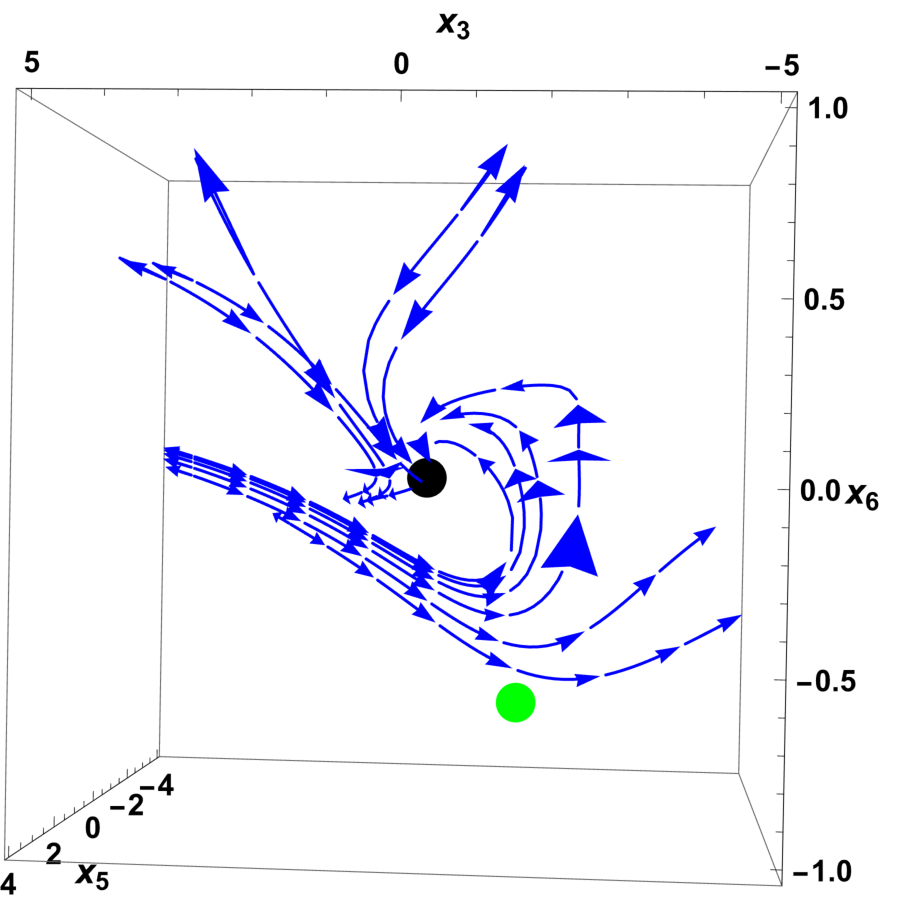}\vspace{2mm}
\includegraphics[width=.45\textwidth]{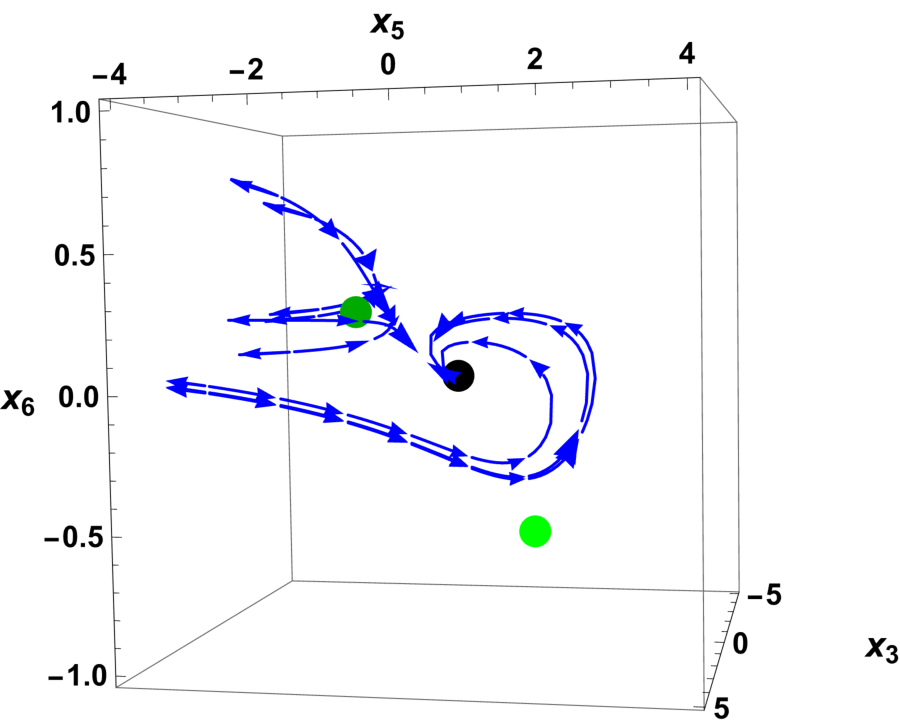}\hspace{2mm}
\includegraphics[width=.4\textwidth]{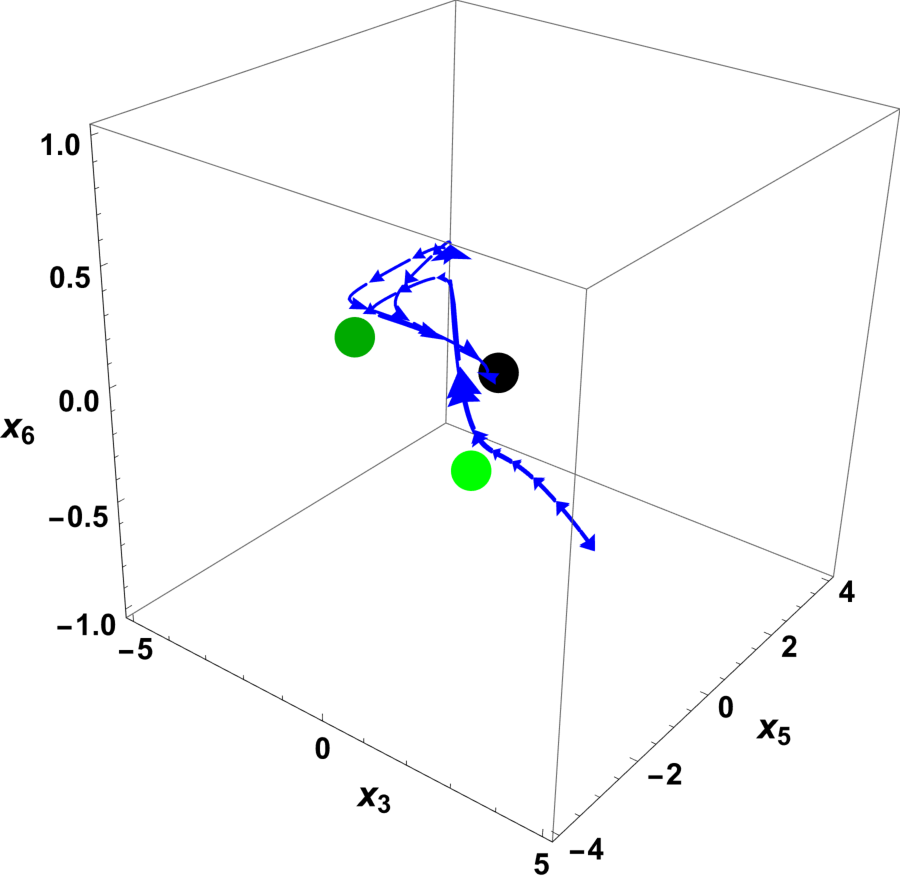}
\caption{Cosmic transitions in $f(Q,C)=\alpha Q^{u}+\beta C^{v}$ gravity with $u=1, v=0.95, r=-3/2$, case of connections type $II$. The green point represents $P^{{\sf PLM}}$, the black point corresponds to $P^{{\sf dS}}$, the darker green point denotes $P^{{\sf X}}$. Upper-Left panel: A $P^{{\sf PLM}} \to P^{{\sf dS}}$ transition with a stable $P^{{\sf dS}}$. Upper-Right panel: the same transition but with an unstable $P^{{\sf dS}}$. Lower panels: a double {\sf PLM} scenario with a stable $P^{{\sf dS}}$, realized for $r=-3/2$. For $u=1/2$, a double {\sf PLM} model can also be constructed; however, it leads to non-acceptable values of the matter density parameter.}\label{fig1}
\end{figure}

\begin{table}[h!]
\centering
\caption{True fixed points for models with $u \neq v$ for connection type $II$.}
\begin{tabular}{l @{\hskip 0.1in} l@{\hskip 0.1in} l @{\hskip 0.1in}l @{\hskip 0.1in}l@{\hskip 0.1in}l@{\hskip 0.1in}l}\hline\hline

Fixed point     &$(x_3,x_5,x_6,x_7,x_8)$           &Stability  &$\Omega_{\sf m}$   &$\Omega_{\sf DE}$      &$w_{eff}$&$(u,v,r)$\\[0.5 ex]
\hline
$P^{{\sf PLM}}$&$\left(-\frac{3}{2},1,-\frac{2}{3},0,0\right)$&Saddle&$1$&$0$&$0$&$ \left(1,-\frac{1}{2},r\right)$\\[0.75 ex]
$P^{{\sf PLM}}$&$\left(-\frac{3}{2},1,-\frac{2}{3},0,0\right)$&Saddle&$1$&$0$&$0$&$ \left(1,v\approx1,r\right)$\\[0.75 ex]
$P^{{\sf dS}}$&$\left(0,0,0,\frac{1-2u}{v-1},0\right)$&Saddle&$0$&$1$&$-1$&$ \left(u,v\neq 1,r\right)$\\[0.75 ex]
$P^{{\sf X}}$&$\left(-\frac{3}{2},-2,-\frac{2}{9v},0,0\right)$&Saddle&$1$&$0$&$0$&$ \left(1,v,-\frac{3}{2}\right)$\\[0.75 ex]

\hline
\end{tabular}
\label{tab1}
\end{table}


\subsubsection{Cosmological Scenarios}\label{type2.1.7}

Next, we examine how the Universe’s evolution is depicted in the proposed model. Crucially, the fixed points must be interconnected for identical values of $u$ and $v$; without such connections, the model fails to describe transitions between cosmological eras in agreement with the established cosmic timeline. Using this criterion, we summarize all types of transitions and their corresponding properties below.

\begin{figure}[h!]
\centering
\includegraphics[width=.5\textwidth]{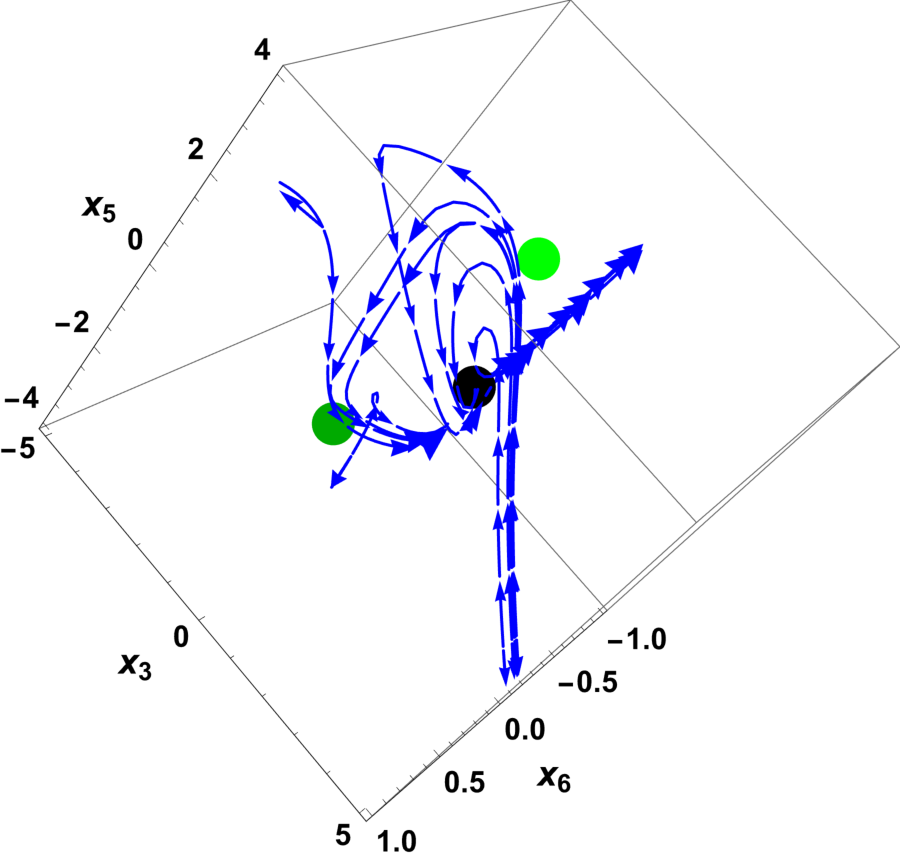}\hspace{2mm}
\includegraphics[width=.45\textwidth]{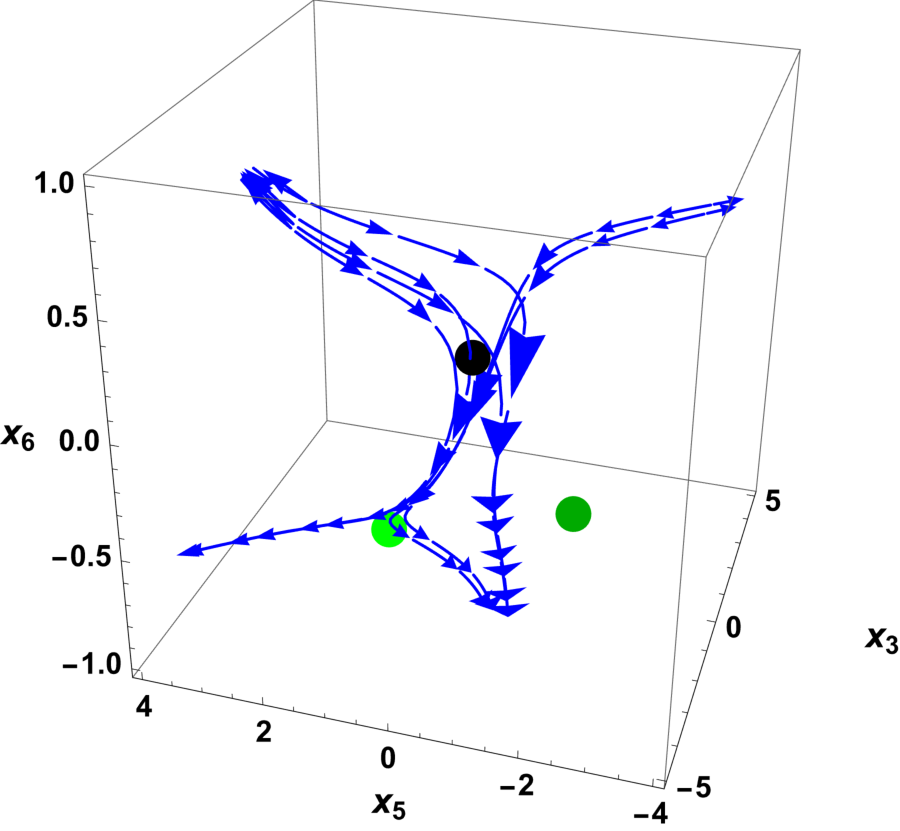}
\caption{Left panel: Illustration showing that in $f(Q,C)=\alpha Q^{u}+\beta C^{v}$ gravity with $u=1, v=0.95$, no early inflationary solution exists. The de Sitter point $P^{{\sf dS}}$ attracts the phase-space trajectories while repelling them away from the {\sf PLM} points. Right panel: For $u=1, v=-1/2$ an early inflationary era connecting to $P^{{\sf PLM}}$ appears.}\label{fig1.1}
\end{figure}

\begin{itemize}

\item $P^{{\sf PLM}}/P^{{\sf X}} \to P^{{\sf dS}}$; A true transition with $\Omega_{{\sf PLM}}=1$ and $\Omega_{{\sf dS}}=1-\Omega_{{\sf m}}|_{P^{{\sf dS}}}=1$ is realized only for $(u=1, v\approx1)$. Here, the de Sitter solution functions as an attractor in the $(x_3,x_5,x_6)$ subspace, though it is unstable in the full phase space. The upper-left panel of Fig.~\ref{fig1} depicts this phase-space behavior, with the green and black points representing $P^{{\sf PLM}}$ and $P^{{\sf dS}}$, respectively.All trajectories that approach $P^{{\sf PLM}}$ ultimately end at $P^{{\sf dS}}$.These models are obtained regardless of the choice of $r$.\\

Scenarios with a saddle de Sitter point as the representative of accelerated expansion in the late times, are not perfect, as the de Sitter solution serves as the final attractor only for certain initial conditions near $P^{{\sf PLM}}$. Besides, for some initial points de Sitter solution is a temporal behavior of system. In the upper-right panel of Fig.~\ref{fig1}, some trajectories are seen to recede from $P^{{\sf dS}}$ after initially approaching it.\\

These models also feature an intriguing property. As discussed in subsection~\ref{type2.1.4}, $P^{{\sf X}}$ can correspond to a {\sf PLM} solution when $x_3=r=-3/2$, representing another {\sf PLM}-dominated state with $\Omega_{{\sf m}}|_{P^{{\sf X}}}^{r=-3/2}=1$ for $u=1$ or $u=1/2$. Thus, for $r=-3/2$ and either choice of $u$, the system exhibits two distinct {\sf PLM} solutions: one from $P^{{\sf PLM}}$ and another from $P^{{\sf X}}$. Their differing locations in phase space correspond to different cosmic times. The lower left panel of Fig.~\ref{fig1} illustrates this scenario, with the darker green point denoting $P^{{\sf X}}$. Two separate sets of trajectories pass through these points, both ultimately reaching $P^{{\sf dS}}$.\\

The scenarios illustrated in Fig.~\ref{fig1} are constructed using a single matter solution ($w_{{\sf eff}}=0$), implicitly assuming that dark matter and baryonic matter dominate simultaneously. In reality, dark matter decouples earlier than baryonic matter, implying that {\sf PLM} could occur twice along the cosmic timeline. This aligns with the existence of two {\sf PLM} fixed points with different origins, a feature naturally incorporated within $f(Q,C)$ gravity theories. \\

\item $P^{{\sf PLM}} \to P^{{\sf X}} \to P^{{\sf dS}}$; Moreover, the Universe can experience a transition belonging to this category, in which {\sf PLM} domination occurs twice prior to the de Sitter phase (see the lower right panel of  Fig.~\ref{fig1}).

\item $P^{{\sf Rad}}/P^{{\sf SM}} \to P^{{\sf PLM}}/P^{{\sf dS}}$; These categories do not admit physically relevant transitions, owing to either incompatible $(u,v)$ values or a vanishing matter density parameter.

\item $P^{{\sf X}} \to P^{{\sf PLM}}/P^{{\sf dS}}$; In the case $r=-2$, $P^{{\sf X}}$ represents a fixed point corresponding to an {\sf URM}-dominated era, with $u=\frac{1}{2} \left(1\pm \sqrt{3}\right)$. None of these configurations allow for a radiation-dominated era that transitions to $P^{{\sf PLM}}$. For $r=-3$, $P^{{\sf X}}$ describes a {\sf SM}-dominated solution with $u=0,5/4$, again showing the same inconsistency.

\item $P^{{\sf dS}}\to P^{{\sf PLM}}/P^{{\sf X}}$ and $P^{{\sf dS}}\to P^{{\sf PLM}}~\&~P^{{\sf X}}$: Although $P^{{\sf dS}}$ is a saddle point, one might consider whether it could act as an initial point from which trajectories move toward $P^{{\sf PLM}}$, representing an early inflationary phase. As shown in the left panel of Fig.~\ref{fig1.1}, trajectories originating near $P^{{\sf dS}}$ do not connect directly to either {\sf PLM} point when $v\approx 1$ is set; trajectories starting further away encounter the {\sf PLM} points first and then proceed to $P^{{\sf dS}}$. Consequently, $P^{{\sf dS}}$ serves as a final attractor, characterizing the late-time dark energy epoch. Nevertheless, in the case $v=-1/2$, as illustrated in the right panel of Fig.~\ref{fig1.1}, point $P^{{\sf dS}}$  (characterized by eigenvalues $(0,-3,3,-9,6)$) represents an early inflationary era to which $P^{{\sf PLM}}$ can provide an effective connection.
\end{itemize}

\subsection{$f(Q,C)=\alpha Q^{u}+\beta C^{v}$ with $u= v$}\label{type2.2}

In this case using the definitions of $x_{1}$, $x_{2}$ and $x_{7}$ from (\ref{t2.5}) one gets

\begin{align}\label{t2.49}
x_7=\frac{\beta }{\alpha } \left(\frac{x_2}{x_1}\right)^v,
\end{align}

which implies that the system is described by a four-dimensional phase space. We choose $(x_{3}, x_{5}, x_{6}, x_{8})$ as the independent variables to construct it. In this case only the equations involving $x_{7}$ are modified. Accordingly, the four-dimensional dynamical system consists of eqs.~(\ref{t2.23}),~(\ref{t2.25}),~(\ref{t2.26}) and~(\ref{t2.28}) with $x_{7}$ determined from (\ref{t2.49}). This system of equations contains the following critical points which also satisfy the conservation of energy momentum tensor~(\ref{t2.2.1}).


\subsubsection{Solution indicating the {\sf PLM}}\label{type2.2.1}

There exists a fixed point corresponding to $w_{{\sf eff}}=0$ , which is characterized by
\begin{align}
&P^{{\sf PLM}}:\left(x_1= -6,x_2=9, x_3= -\frac{3}{2}, x_4=0, x_5= 1, x_6=-\frac{2}{3}, x_8= 0\right)\label{t2.50}\\
&\mbox{Ex.}\to \beta=\frac{\left(-\frac{3}{2}\right)^{-v}(1 -2 v)\alpha }{(v-1) (2 v+1)},~\mbox{Ei.}=\left[-3v,R_{1},R_{2},R_{3}\right],~R_{i}=\frac{\text{Root}[\mathfrak{F}(v)\&,i]}{6(2 v-1)},~~\Omega_{{\sf m}}|_{P^{{\sf PLM}}}=0,\label{t2.51}\\
&\mathfrak{F}(v)=\#^3-6 (2 v-1) (6 v+7)\#^2+27(2 v-1) \Big[2 v (80 v-11)-41\Big] \# -486 (1-2 v)^2 \Big[4 v \left(6 v^2+34 v-33\right)+41\Big],\nonumber
\end{align}

where $R_{i}$ denotes the $i$-th root of $\mathfrak{F}(v)$. As can be seen, this solution is not a genuine one since it fails to represent a {\sf PLM}-dominated epoch. In fact, it corresponds to a non-effective matter era. Therefore, we discard this critical point as non-physical.


\subsubsection{de Sitter solution}\label{type2.2.2}

A de Sitter solution also exists in this case with the following aspects

\begin{align}
&P^{{\sf dS}}:\left(x_1= -6,x_2=18, x_3= 0, x_4=0, x_5=0, x_6=0, x_8= 0\right)\label{t2.52}\\
&\mbox{Ex.}\to \beta=\frac{3^{-v} e^{-i \pi  v} (1 -2 v)\alpha}{v-1},~\mbox{Ei.}=\left[0,R_{1},R_{2},R_{3}\right],~R_{i}=\frac{\text{Root}[\mathfrak{G}(v)\&,i]}{6(1-2 v)},~~\Omega_{{\sf DE}}|_{P^{{\sf dS}}}=1,\label{t2.53}\\
&\mathfrak{G}(v)=\#^3-6 e^{i \pi  v} (2 v-1)\#^2 +324 e^{2 i \pi  v} (2 v-1) (6 v-5) \#+23328 e^{3 i \pi  v} (v-1) (1-2 v)^2.\nonumber
\end{align}

The relation between $\alpha$ and $\beta$ given in~(\ref{t2.52}) yields real values only when $v\in \mathbb{Z}$. Hence, the de Sitter solution exists exclusively for integer values of $v$. Although analyzing the stability regions is not straightforward, we will later demonstrate the stability of $P^{{\sf dS}}$ for specific values $v$.


\subsubsection{Solutions with $w_{{\sf eff}}=-1-\frac{2}{3}r$}\label{type2.2.3}

Since $P^{{\sf PLM}}$ provide a genuine solution corresponding to the {\sf PLM}, we turn to solutions of the system for $x_{3}=r=-3/2$. In this case, the function $\gamma=\gamma_{0}a^{r}$ is responsible for providing this behavior. is responsible for realizing this behavior. In fact, three such critical points arise, as follows.

\begin{align}
&P^{{\sf Y}}_{1}:\left(x_1= 3, x_2=0, x_3= -\frac{3}{2}, x_4=2, x_5= -2, x_6=\frac{2}{9}, x_8= 0\right),\label{t2.54}\\
&\mbox{Ex.}\to (v\gtrsim1,\alpha,\beta),~\mbox{Ei.}=\left[-3,-\frac{9}{2},\frac{1}{4} \left(20-4 \sqrt{14}i \right),\frac{1}{4} \left(20+4 \sqrt{14}i \right)\right],~\Omega_{{\sf m}}|_{P^{{\sf Y}}_{1}}^{v\gtrsim1}\lesssim1.\label{t2.55}
\end{align}

That is $P^{{\sf Y}}_{1}$ is a true solution. In the case $v\gtrsim1$ one obtains $x_{3}\lesssim-\frac{3}{2}$ (equivalently $w_{{\sf eff}}\gtrsim 0$). 

Another solution is given by

\begin{align}
&P^{{\sf Y}}_{2}:\left(x_1= 0, x_2=3, x_3= -\frac{3}{2}, x_4=\frac{4}{3}, x_5= \infty, x_6=-2, x_8= 0\right),\label{t2.60}\\
&\mbox{Ex.}\to \left[v=\frac{1}{12} \left(7-\sqrt{73}\right)\approx-0.13,\alpha,\beta\right],~\mbox{Ei.}=\left[\frac{25}{2},-\frac{\sqrt{73}}{2},-3,\frac{1}{4} \left(\sqrt{73}-7\right)\right],~\Omega_{{\sf m}}|_{P^{{\sf Y}}_{4}}^{v=-0.13}=\frac{1}{6} \left(\sqrt{73}+11\right)\approx 3.25,\label{t2.61}\\
&\mbox{Ex.}\to \left[v=\frac{1}{12} \left(7+\sqrt{73}\right)\approx 1.3,\alpha,\beta\right],~\mbox{Ei.}=\left[\frac{25}{2},\frac{\sqrt{73}}{2},-3,-\frac{1}{4} \left(\sqrt{73}+7\right)\right],~\Omega_{{\sf m}}|_{P^{{\sf Y}}_{4}}^{v=1.3}=\frac{1}{6} \left(\sqrt{73}-11\right)\approx 0.41.\label{t2.62}
\end{align}

Point~(\ref{t2.62}) is regarded as a solution located at infinity. Similar to solution~(\ref{t2.61}), there also exists another solution which cannot be considered as physically valid, since at this point the matter density parameter exceed the admissible range $0<\Omega_{{\sf m}}<1$. Its characteristics are given by

\begin{align}
&P^{{\sf Y}}_{3}:\left(x_1= 3,x_2=0, x_3= -\frac{3}{2}, x_4=2, x_5= -2, x_6=\frac{4}{9}, x_8= 0\right),\label{t2.58}\\
&\mbox{Ex.}\to \left(v=\frac{1}{2},\alpha,\beta\right),~\mbox{Ei.}=\left[-\frac{3}{2},-\frac{9}{2},\frac{1}{9} \left(45-9 i \sqrt{14}\right),\frac{1}{9} \left(45+9 i \sqrt{14}\right)\right],~\Omega_{{\sf m}}|_{P^{{\sf Y}}_{3}}^{v=\frac{1}{2}}=3\label{t2.59}
\end{align}

Hence, $P^{{\sf Y}}_{1}$ and $P^{{\sf Y}}_{2}$ specified with~(\ref{t2.62}) represent two true {\sf PLM}-dominated solutions. However, we select $P^{{\sf Y}}_{1}$ as the only true solution, since, the observations confirms that the Universe could not have evolved from an initial {\sf PLM}-dominated state.


\subsubsection{{\sf URM} solutions}\label{type2.2.4}

In the case $u=v$, {\sf URM} emerges as a solution of the system for

\begin{align}
&P^{{\sf Rad}}:\left(x_1= -6,x_2=6, x_3= -2, x_4=0, x_5= \frac{4}{3}, x_6=-\frac{4}{3}, x_8= 0\right),\label{t2.63}\\
&\mbox{Ex.}\to \beta \to \frac{\alpha  (-1)^{1-v} (2 v-1)}{(v-1) (4 v+1)},~\mbox{Ei.}=\left[-3v,R_{1},R_{2},R_{3}\right],~R_{i}=\frac{\text{Root}[\mathfrak{I}(v)\&, i]}{18(-1+2v)},~~\Omega_{{\sf m}}|_{P^{{\sf Rad}}}=0,\label{t2.64}\\
&\mathfrak{I}(v)=\#^3-18  (2 v-1) (8 v+9)\#^2+324 (2 v-1) \Big[2 v (96 v-23)-31\Big]\#^2-11664 (1-2 v)^2 \Big[8 v \big(v (4 v+41)-34\big)+59\Big],\nonumber
\end{align}

which apparently is not a true solution.


\subsubsection{{\sf SM} solutions}\label{type2.2.5}

In this case the equations give

\begin{align}
&P^{{\sf SM}}:\left(x_1= -6,x_2=0, x_3= -3, x_4=0, x_5= 2, x_6=\frac{1}{3 v}, x_8= 0\right),\label{t2.65}\\
&\mbox{Ex.}\to (v, \alpha, \beta),~\mbox{Ei.}=\left[8+i \sqrt{23},8-i \sqrt{23},-3,-6 v\right],~\Omega_{{\sf m}}|_{P^{{\sf SM}}}=2-\frac{1}{v}.\label{t2.65}
\end{align}

An interesting point is that $P^{{\sf SM}}$ constitutes a true fixed point for $\frac{1}{2}<v\leq 1$, particularly at $v=1$, where one finds $\Omega_{{\sf m}}|_{P^{{\sf SM}}}=1$. Table~\ref{tab2} provides a compact overview of the viable cosmological solutions obtained for connection type $II$ for $u=v$.

\begin{table}[h!]
\centering
\caption{True fixed point solutions for models with $u=v$ in the case of connections type $II$.}
\begin{tabular}{l @{\hskip 0.1in} l@{\hskip 0.1in} l @{\hskip 0.1in}l @{\hskip 0.1in}l@{\hskip 0.1in}l@{\hskip 0.1in}l}\hline\hline

Fixed point     &$(x_3,x_5,x_6,x_8)$           &Stability  &$\Omega_{\sf m}$   &$\Omega_{\sf DE}$      &$w_{eff}$&$(v,r,\alpha,\beta)$\\[0.5 ex]
\hline

$P^{{\sf dS}}$&$\left(0,0,0,0\right)$&depends on $v$ &$0$&$1$&$-1$&$ \left(v\in \mathbb{Z},r,\alpha,\frac{3^{-v} e^{-i \pi  v} (1 -2 v)}{v-1}\alpha\right)$\\[0.75 ex]
$P^{{\sf Y}}_{1}$&$\left(r,\frac{2 r}{3 (r+2)},\frac{-2 v^2+v+3}{15 v-6 v^3},0\right)$&Saddle&$\lesssim1$&$0$&$0$&$ \left(v\gtrsim1,\frac{3 (2-v)}{(v+1) (2 v-3)},\alpha,\beta,\right)$\\[0.75 ex]
$P^{{\sf SM}}$&$\left(-3,2,\frac{1}{3v},0\right)$&Saddle&$2-\frac{1}{v}$&$0$&$0$&$ \left(v\in(\frac{1}{2},1],r,\alpha,\beta\right)$\\[0.75 ex]

\hline
\end{tabular}
\label{tab2}
\end{table}

\subsubsection{Cosmological Scenarios}\label{type2.2.6}

In this case, when $u=v$, $P^{{\sf PLM}}$ does not act as a {\sf PLM}-dominated solution and $P^{{\sf Y}}_{1}$ serves this role for $v=u\gtrsim1$. Under the same assumptions, the de Sitter and {\sf SM} solutions are accounted for true fixed points. Consequently, all cosmological scenarios are constructed from interconnections among these three points in phase space. Importantly, the de Sitter solution is the only one corresponding to an accelerated expansion phase, and it represents an unstable fixed point. This implies that $P^{{\sf dS}}$ can play a role in scenarios involving either early- or late-time inflation. We briefly discuss some of these cases below.

\begin{figure}[h!]
\centering
\includegraphics[width=.45\textwidth]{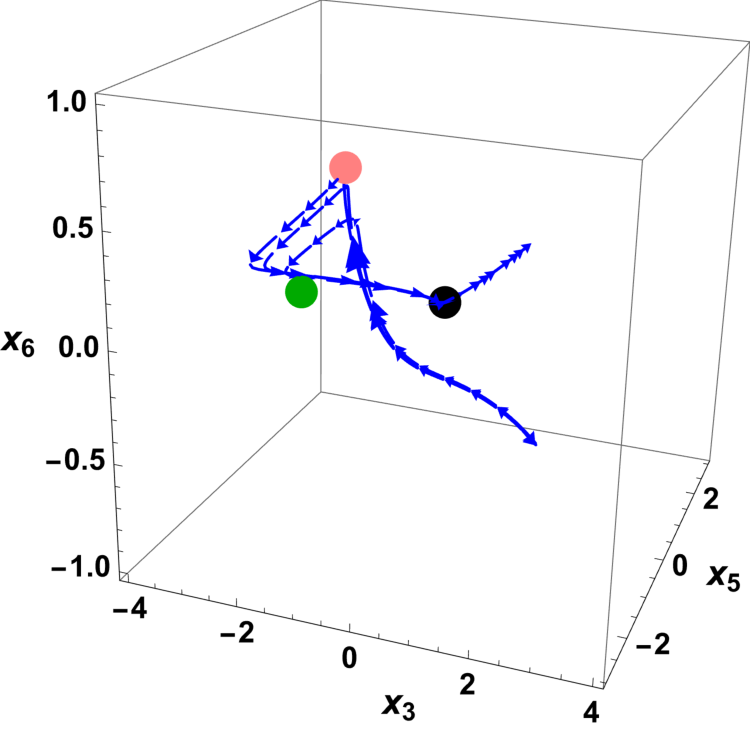}\hspace{2mm}
\includegraphics[width=.45\textwidth]{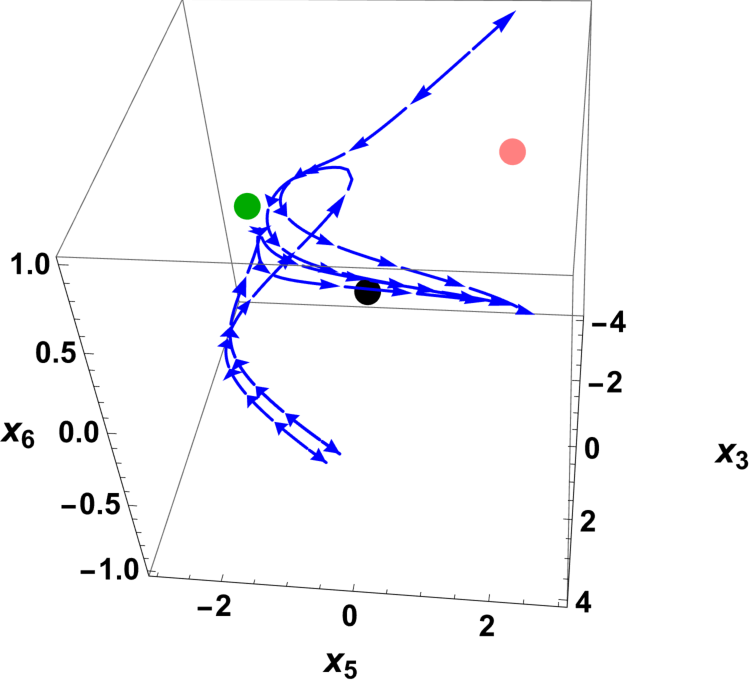}\vspace{2mm}
\includegraphics[width=.45\textwidth]{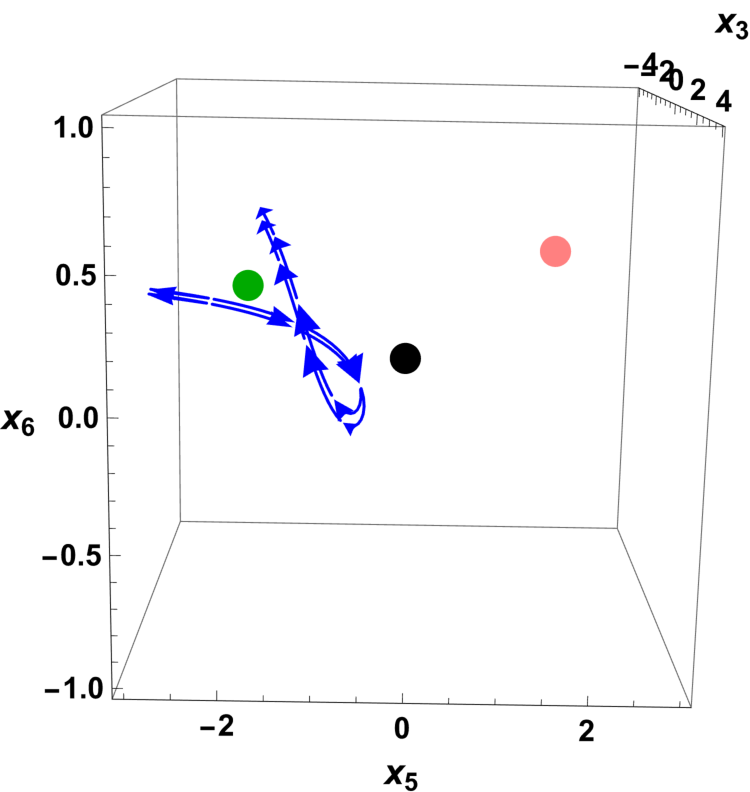}\hspace{2mm}
\includegraphics[width=.45\textwidth]{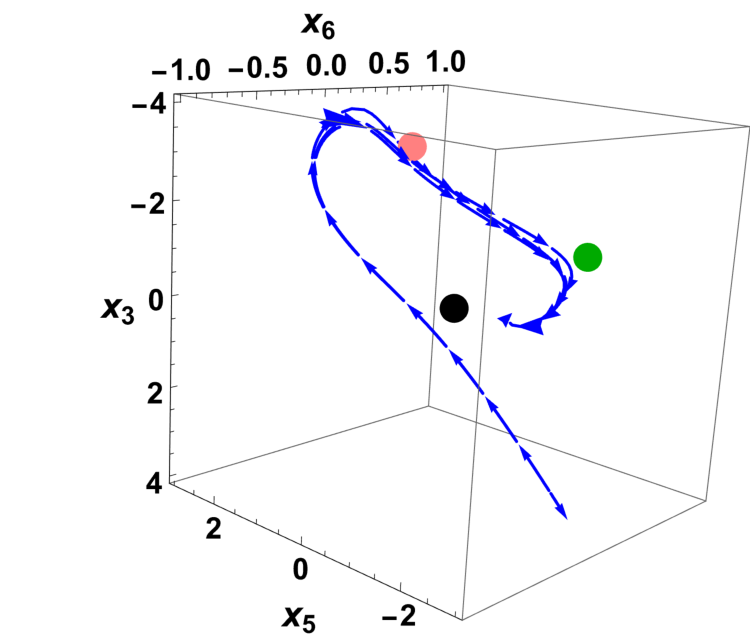}
\caption{Different scenarios in the case $u=v=1.1$ using connections type $II$. These panels illustrate models with a single {\sf PLM}-dominated era. The de Sitter solution $P^{{\sf dS}}$ represents inflation at both early times (lower panel) and late times (upper panel).}\label{fig2}
\end{figure}

\begin{itemize}

\item Transitions of type $P^{{\sf SM}} \to P^{{\sf Y}}_{1} \to P^{{\sf dS}}$, as illustrated in the upper-left panel of Fig.~\ref{fig2}. In Fig.~\ref{fig2} the pink, the dark green and the black points indicate $P^{{\sf SM}}$, $P^{{\sf Y}}_{1}$ and $P^{{\sf dS}}$, respectively. Such an evolution could hypothetically occur in theories where preheating proceeds with a delay.

\item There is also a $P^{{\sf Y}}_{1}\to P^{{\sf dS}}$, as depicted in the upper-right panel of Fig.~\ref{fig2}.

\item The $P^{{\sf dS}}\to P^{{\sf Y}}_{1}$ transitions are shown in the lower-left panel of Fig.~\ref{fig2}.

\item The transition $P^{{\sf dS}}\to P^{{\sf SM}} \to P^{{\sf Y}}_{1} \to P^{{\sf dS}}$ represents a period of stiffness between the early inflationary era and dark matter production, as predicted by some theories. The lower-right panel of Fig.~\ref{fig2} illustrates this scenario. In this figure, the Universe evolves from a de Sitter era and ultimately returns to a de Sitter era, both represented by $P^{{\sf dS}}$.

\end{itemize}

\section{Connection type $III$}\label{type3}

For connection type $III$, the equations of motion take the form

\begin{align}
&\kappa \rho_m=\frac{f}{2}+f_Q \left(3 H^2-\frac{Q}{2}\right)-\frac{C f_C}{2}-\frac{3 \gamma  \dot{f}_Q}{2a^{2}}+\left(3 H+\frac{3 \gamma}{2a^{2}}\right)\dot{f}_C,\label{t3.1}\\
&\kappa p_m=-\frac{f}{2}+f_Q \left(-3 H^2-2 \dot{H}+\frac{Q}{2}\right)+\frac{C f_C}{2}+\frac{1}{2}\left (\frac{ \gamma}{a^{2}} -4 H\right) \dot{f}_Q-\frac{ \gamma  \dot{f}_C}{2a^{2}}-\ddot{f}_C,\label{t3.2}\\
&\kappa\m\nabla_{\mu}T^{\mu}_{t}=\mathcal{C}_{III},\label{t2.2.1}\\
&\mathcal{C}_{III}=-\frac{3\gamma}{2 a^2}\left[\left(H+2\frac{\dot{\gamma}}{\gamma}\right)\dot{f}_Q+\ddot{f}_Q-\left(H+2\frac{\dot{\gamma}}{\gamma}\right)\dot{f}_C-\ddot{f}_C\right],
\end{align}

where the definitions of $Q$ and $C$are modified slightly, as follows

\begin{align}
&Q=-6 H^2+\frac{3}{a^{2}}\left( \gamma  H+ \dot{\gamma}\right),\label{t3.3}\\
&C=6 \left(3 H^2+\dot{H}\right)-\frac{3}{a^{2}}\left( \gamma  H+ \dot{\gamma}\right),\label{t3.4}\\
\end{align}

and the relation~(\ref{t2.4.1}) remains unchanged. We note that the scale factor $a$ explicitly appears in the equations of motion for a type $III$ connection. Using the same definitions as~(\ref{t2.5}) and~(\ref{t2.6}), except for $x_{4}$, one obtains 

\begin{align}
&\Omega_{{\sf m}} =1-\Omega_{{\sf DE}},\nonumber\\
&\Omega_{{\sf DE}} =\frac{1}{12 j}\Bigg\{x_1 \bigg[j \left(3 d x_4 x_5+2\right)-x_7 \Big(3 e i \left(x_4+2\right) x_6+2 (i+1)\Big)-2\bigg]\Bigg\},\label{t3.5}\\
&4\frac{\ddot{C}}{H^{4}}=-\frac{{x_2}^{2}}{ e i x_1 x_7}\Bigg\{j \bigg[-x_1 \Big(d \left(x_4-4\right) x_5+2\Big)+8 x_3+12\bigg]+\nonumber\\
& x_1 \bigg[x_7 \Big(i \big(e x_6 \left(p x_2 x_6+x_4\right)-2\big)+2\Big)+2\bigg]\Bigg\},\label{t3.6}\\
&4\frac{\ddot{Q}}{H^{4}}=12 x_1 \left(2 x_3-3\right)-24 \big[2 \left(x_3-10\right) x_3+9\big]+\\
&12 x_4 \big[m n \left(l+3 x_3+3\right)+n \left(1-x_3\right)-39 x_3+27\big]+x_5 x_1^2 \big[(n+3) x_4+2\big]+x_2^2 x_6 \big[(n+3) x_4-4\big],\label{t3.6.1}\\
&\mathcal{C}_{III}=-\frac{3}{8} H x_1 x_4 x_8 \left\{u(u-1) \left[\frac{4 \ddot{Q}}{H^4 x_1^2}+x_5 \left(4 r+(u-2) x_1 x_5+2\right)\right]+v(1-v)  x_7 \left[\frac{4 \ddot{C}}{H^4 x_2^2}+x_6 \left(4 r+(v-2) x_2 x_6+2\right)\right]\right\},\label{t3.6.2}
\end{align}

for eqs.~(\ref{t3.1}),~(\ref{t3.2}), the expression of $\ddot{Q}$ for connection type $III$ and $\mathcal{C}_{III}$. In eqs.~(\ref{t3.5}) and~(\ref{t3.6}) we redefine $x_{4}=\gamma/(Ha^{2})$. Relation~(\ref{t2.9}) remains unchanged; however, the counterparts of~(\ref{t2.17}) and~(\ref{t2.18}) are now given by

\begin{align}
&3 (n+1) x_4+x_2=6 \left(x_3+3\right),\label{t3.7}\\
&2 x_4\big[n \left(m+x_3\right)-2\big]+2 x_3 x_4=3 x_5 \big[(n+1) x_4-2\big]^2+8 x_3\label{t3.8}.
\end{align}

By proceeding in a manner similar to that described after relation~(\ref{t2.9}) and noting that eqs.~(\ref{t2.14})-(\ref{t2.16}) still hold we reach at a set of equations in which the same equations as~(\ref{t2.21})-(\ref{t2.24}) and~(\ref{t2.27})-(\ref{t2.28}) appear with some modifications in~(\ref{t2.25}) and (\ref{t2.26}) as follows

\begin{align}
&\frac{dx_{5}}{dN}|_{III}=\frac{dx_{5}}{dN}|_{II}+\frac{72-\left(x_4+6\right) x_5 x_1^2-12 \left(x_3-1\right) x_1-156 x_4-24 x_3 \left(9-8 x_4\right)-x_2^2 x_4 x_6}{x_1^2},\label{t3.9}\\
&\frac{dx_{6}}{dN}|_{III}=\frac{dx_{6}}{dN}|_{II}+x_4 \left(x_6-\frac{d j x_5}{e i x_7}\right)\label{t3.10},
\end{align}

where the subscripts $II$ and $III$ indicate equations obtained by considering connections type $II$ and $III$, respectively.


\subsection{$f(Q,C)=\alpha Q^{u}+\beta C^{v}$ with $u\neq v$}\label{type3.1}

Using the definitions~(\ref{t2.6}) and assuming power-law forms for the functions, as discussed in the previous section, the resulting system of equations (as explained after relation~(\ref{t3.8})) contains critical points, which are described below. 


\subsubsection{{\sf PLM} solutions}\label{type3.1.1}

There are two true solutions that describe {\sf PLM} domination in the evolution of the Universe. These solutions occupy different locations in phase space, a feature also observed in the previous section. Additionally, there exists a solution that is independent of the choice of the $\gamma$, and it has the following properties

\begin{align}\label{t3.11}
&P^{{\sf PLM}}_{1}:\left(x_1=-6, x_2=9, x_3= -\frac{3}{2}, x_4=0, x_5= 1, x_6= -\frac{2}{3}, x_7=0, x_8= 0\right)\nonumber\\
&\mbox{Ex.}\to \left(u=1, v=-\frac{1}{2}\right),~\mbox{Ei.}=\left[-3, \frac{9}{2},\frac{15}{2},-\frac{1}{4} \left(\sqrt{\frac{96}{\text{x7}}+9}+9\right),\frac{1}{4} \left(\sqrt{\frac{96}{\text{x7}}+9}-9\right)\right],~\Omega_{{\sf m}}|_{P^{{\sf PLM}}_{1}}=1.
\end{align}

Another solution exists only for a specific value of $r$ (corresponding to a particular $\gamma$ function) exhibits the following properties

\begin{align}\label{t3.12}
&P^{{\sf PLM}}_{2}:\left(x_1=3, x_2=0, x_3= -\frac{3}{2}, x_4=2, x_5= 0, x_6= \frac{2}{5 v}, x_7=0, x_8= 0\right)\nonumber\\
&\mbox{Ex.}\to \left(u=1, r=\frac{1}{2}\right),~\mbox{Ei.}=\left[0,-3,-\frac{5}{2},-2 \sqrt{3}-\frac{3}{2},2 \sqrt{3}-\frac{3}{2}\right],~\Omega_{{\sf m}}|_{P^{{\sf PLM}}_{2}}=1.
\end{align}

We observe that both fixed points are unstable and exist only when $u=1$ is chosen.


\subsubsection{de Sitter solution}\label{type3.1.2}

Models constructed using connection type $III$ contain a de Sitter solution at the same location as mentioned in~(\ref{t2.35}), which is always unstable for $(u=1, v=-\frac{1}{2})$ and stable for $(u=1, r=\frac{1}{2})$ when $0<v<1$. We denote this solution as $P^{{\sf dS}}_{1}$.Additionally, there are three other de Sitter solutions that exist for $r=2$, with properties given by

\begin{align}\label{t3.13}
&P^{{\sf dS}}_{2}:\left(x_1=12, x_2=0, x_3= 0, x_4=2, x_5=0, x_6=0, x_7= \frac{2-u}{2 (v-1)},x_8\right)\nonumber\\
&\mbox{Ex.}\to r=2,~\mbox{Ei.}=\left[0,0,-1,-3,-3\right],~~\Omega_{{\sf DE}}|_{P^{{\sf dS}}_{2}}=1,\\\nonumber\\
&P^{{\sf dS}}_{3}:\Bigg(x_1=\frac{6 u}{u+(v-1) x_{7}-1}, x_2=12-\frac{6 u}{u+(v-1) x_{7}-1}, x_3= 0, x_{4}=\frac{2 \left(\frac{u}{u+(v-1) x_{7}-1}+1\right)}{r+1}, \\
&x_5=\frac{r-2}{3 u^2}\big[u+(v-1) x_{7}-1) (2 u+(v-1) x_{7}-1\big], x_6=0,x_7,x_8\Bigg)\nonumber\\
&\mbox{Ex.}\to r=2,~\mbox{Ei.}=\mbox{unstable at the location of } P^{{\sf PLM}}_{1},~~\Omega_{{\sf DE}}|_{P^{{\sf dS}}_{3}}=1,\\\nonumber\\
&P^{{\sf dS}}_{4}:\left(x_1=12, x_2=0, x_3= 0, x_4=\frac{6}{r+1}, x_5=\frac{1}{4}(r-2), x_6=\frac{1}{v}, x_7, 0\right)\nonumber\\
&\mbox{Ex.}\to(u=2, r=2),~\mbox{Ei.}=\left[0,0,-1,-3,-3\right],~~\Omega_{{\sf DE}}|_{P^{{\sf dS}}_{4}}=1.
\end{align}

As observed, $P^{{\sf dS}}_{2}$ and $P^{{\sf dS}}_{4}$ are stable whereas $P^{{\sf dS}}_{3}$ is not. Also, note that $P^{{\sf dS}}_{2}$ and $P^{{\sf dS}}_{3}$ have different values of $x_{1}$ and $x_{2}$, which are obtained utilizing relations~(\ref{t2.9}),~(\ref{t2.27}) and~(\ref{t2.28}). Additionally, we have $\mathcal{C}_{III}=0$ for arbitrary values of $x_{8}$ when $r=2$ for $P^{{\sf dS}}_{2}$ and $P^{{\sf dS}}_{3}$ while in case of $P^{{\sf dS}}_{4}$ it must be $x_{8}=0$ even if $r=2$.


\subsubsection{Solutions with $w_{{\sf eff}}=-1-\frac{2}{3}r$}\label{type3.1.3}

For possible dark-energy-dominated solutions, two fixed points exist, which are given by

For possible dark energy dominated solution there exist two fixed points which are found as

\begin{align}
&P^{{\sf Z}}:\left(x_1=-6, x_2=6 (r+3), x_3= r, x_{4}=0, x_5=-\frac{2r}{3}, x_6= \frac{2 r}{3 (r+3)}, x_7=0, x_8= 0\right)\label{t3.14}\\
&\mbox{Ex.}\to \left(u= \frac{1}{2},v=\frac{r+3}{2 r}\right),~\mbox{Ei.}=\left[3,r,-7 r-3,\frac{1}{2} \left(-5 r+\sqrt{r (17 r+24)}-12\right),-\frac{1}{2} \left(5 r+\sqrt{r (17 r+24)}+12\right)\right],\label{t3.15}\\
&\mbox{Ex.}\to \left(u= \frac{1}{2},v= -\frac{3}{2 r}\right),~\mbox{Ei.}=\left[r,-r-3,-7 r-3,\frac{1}{2} \left(\sqrt{3} \sqrt{r (11 r+48)+48}-3 r\right),-\frac{1}{2} \left(3 r+\sqrt{3} \sqrt{r (11 r+48)+48}\right)\right],\label{t3.16}\\
&\Omega_{{\sf DE}}|^{u= \frac{1}{2}}_{P^{{\sf Z}}}=1.\nonumber\\\nonumber\\
&\mbox{Ex.}\to \left(u= -\frac{3}{2 r},v= \frac{r+3}{2 r}\right),~\mbox{Ei.}=\mbox{always unstable},~~~\Omega_{{\sf DE}}|^{u= -\frac{3}{2 r}}_{P^{{\sf Z}}}=\frac{2 (r+3)}{3}.\label{t3.17}
\end{align}

For the solution~(\ref{t3.14}) under the conditions~(\ref{t3.17}): i) it is impossible for the two requirements $r > -1$, which gives $w_{{\sf eff}}<-1/3$ and $0<{2(r+3)}/3<1$, which ensures $0<\Omega_{{\sf DE}}<1$ hold simultaneously. This implies that it is not a true solution. ii) in the case $r=-3/2$ we recover $P^{{\sf PLM}}_{1}$. Therefore, only theories with $u=1/2$ contain solutions with the property $x_{3}=r$, and consequently, connections to $P^{{\sf PLM}}_{1}$ and $P^{{\sf PLM}}_{2}$ in phase space cannot be established. Additionally, critical points corresponding to {\sf URM} domination can be obtained by choosing suitable values of $r$ for $u=1/2, 3/4$, and {\sf SM} domination occurs for $u=1/2$, with properties similar to those discussed in sections~\ref{type2.1.5} and~\ref{type2.1.6}.

Table~\ref{tab3} summarizes the admissible true solutions for the case of connection type $III$.

\begin{table}[h!]
\centering
\caption{True fixed points for models with $u\neq v$ for connection type $III$.}
\begin{tabular}{l @{\hskip 0.1in} l@{\hskip 0.1in} l @{\hskip 0.1in}l @{\hskip 0.1in}l@{\hskip 0.1in}l@{\hskip 0.1in}l}\hline\hline

Fixed point     &$(x_3,x_5,x_6,x_7,x_8)$           &Stability  &$\Omega_{\sf m}$   &$\Omega_{\sf DE}$      &$w_{eff}$&$(u,v,r)$\\[0.5 ex]
\hline

$P^{{\sf PLM}}_{1}$&$\left(-\frac{3}{2},1,-\frac{2}{3},0,0\right)$&Saddle&$1$&$0$&$0$&$ \left(1,-\frac{1}{2},r\right)$\\[0.75 ex]
$P^{{\sf PLM}}_{2}$&$\left(-\frac{3}{2},0,\frac{2}{5v},0,0\right)$&Saddle&$1$&$0$&$0$&$ \left(1,v,\frac{1}{2}\right)$\\[0.75 ex]
$P^{{\sf dS}}_{1}$&$\left(0,0,0,\frac{1-2u}{v-1},0\right)$&depends on $v$ &$0$&$1$&$-1$&$ \left(u,v\neq 1,r\right)$\\[0.75 ex]
$P^{{\sf dS}}_{2}$&$\left(0,0,0,\frac{2-u}{2(v-1)},x_{8}\right)$&Stable &$0$&$1$&$-1$&$ \left(u,v\neq 1,2\right)$\\[0.75 ex]
$P^{{\sf dS}}_{3}$&$\left(0,0,0,x_{7},x_{8}\right)$&unstable&$0$&$1$&$-1$&$ \left(u,v,2\right)$\\[0.75 ex]

\hline
\end{tabular}
\label{tab3}
\end{table}


\subsubsection{Cosmological Scenarios}\label{type3.1.4}

Based on the discussion above, only scenarios involving critical points with the same values $u$ and $v$ can be constructed. In other words, only these fixed points can be connected in phase space. Consequently, the possible evolutionary scenarios for models with connection type $III$ are listed below.

\begin{figure}[h!]
\centering
\includegraphics[width=.45\textwidth]{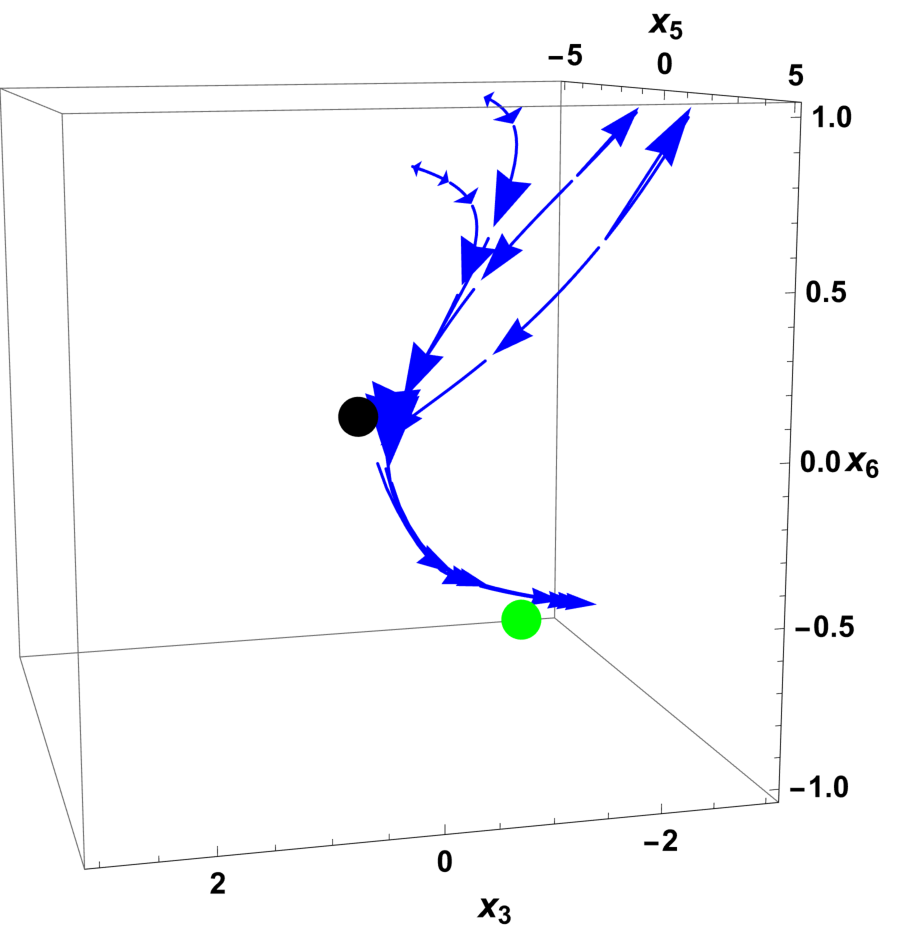}\hspace{2mm}
\includegraphics[width=.45\textwidth]{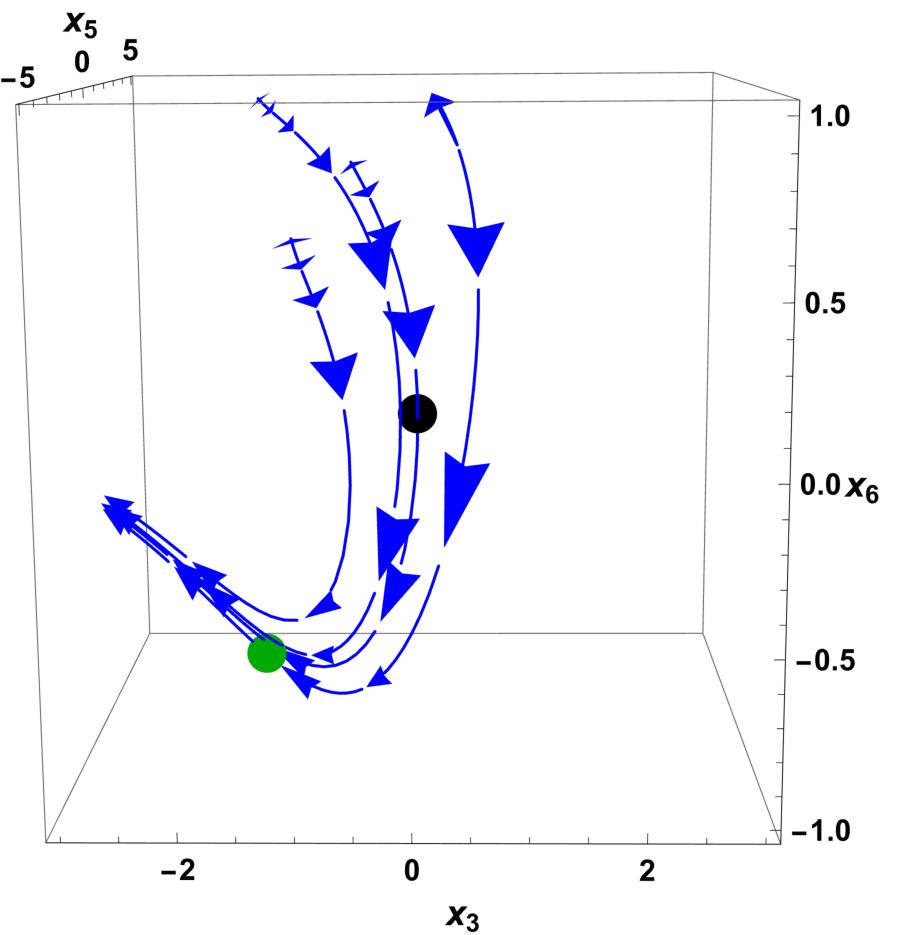}\vspace{2mm}
\includegraphics[width=.45\textwidth]{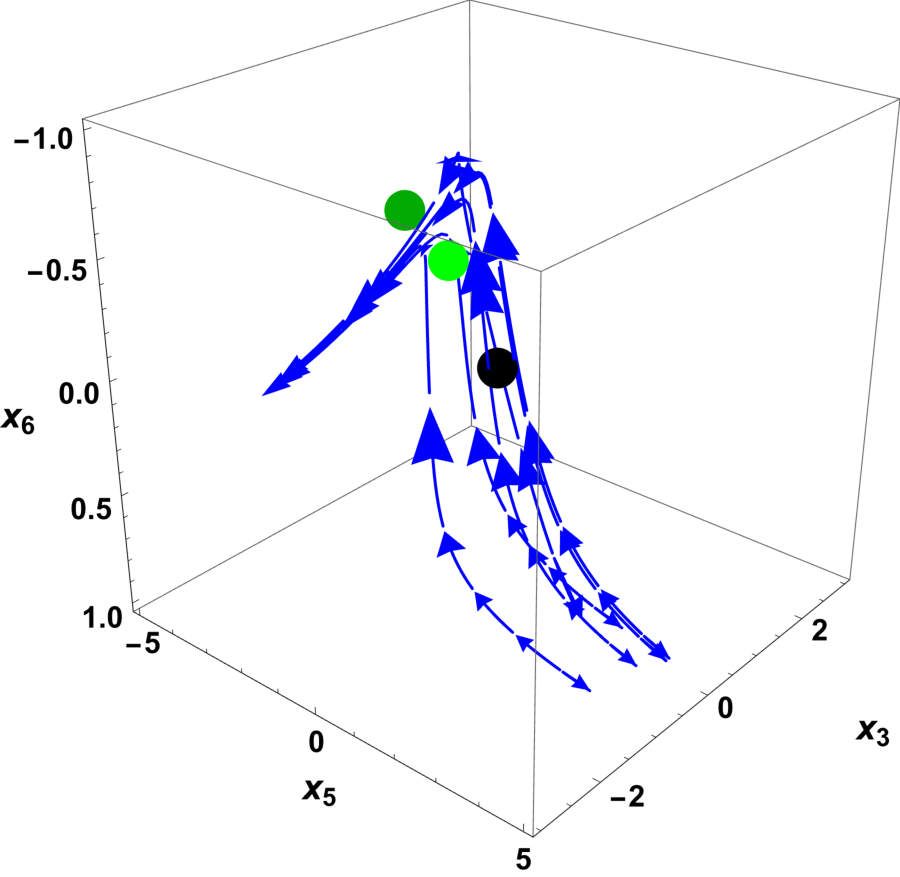}\hspace{2mm}
\includegraphics[width=.45\textwidth]{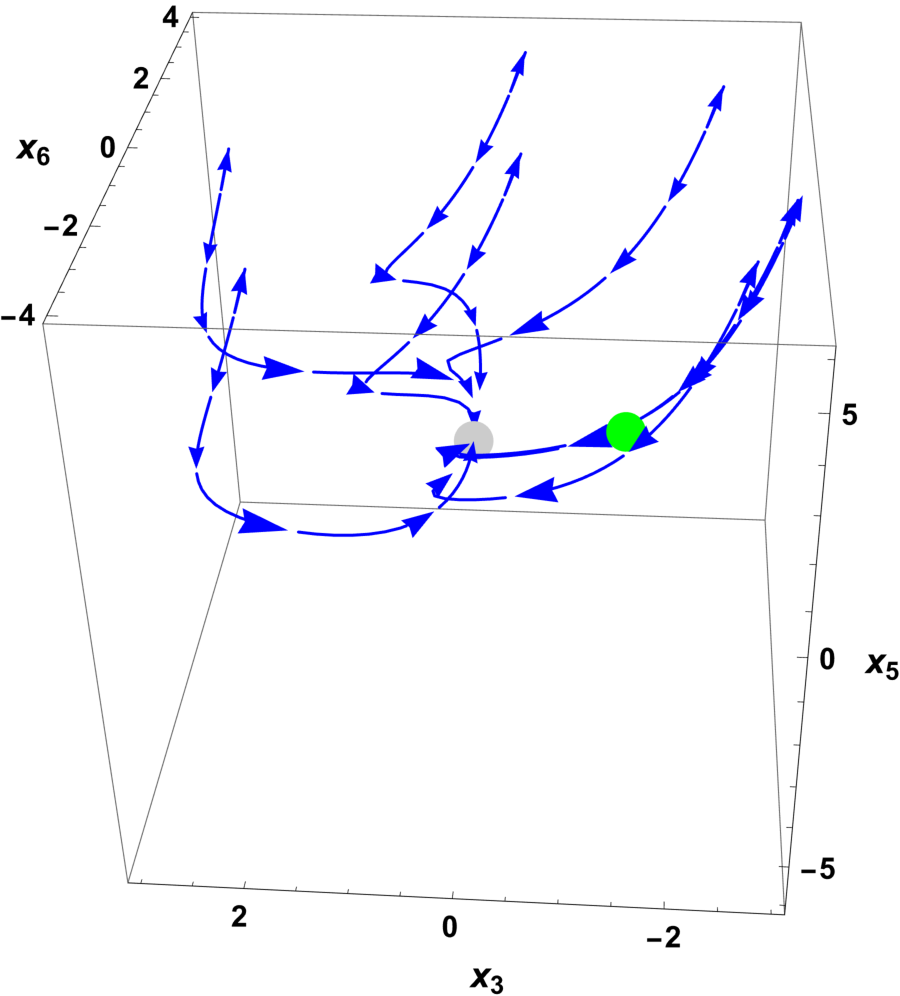}
\caption{Cosmological scenarios obtained for connections type $III$ in the case $u=1, v=-1/2$ and $r=1/2$. In these scenarios $P^{{\sf dS}}_{1}$ remarks an early inflation and $P^{{\sf dS}}_{2}$ provides the late time accelerated expansion.}\label{fig4}
\end{figure}

\begin{figure}[h!]
\centering
\includegraphics[width=.45\textwidth]{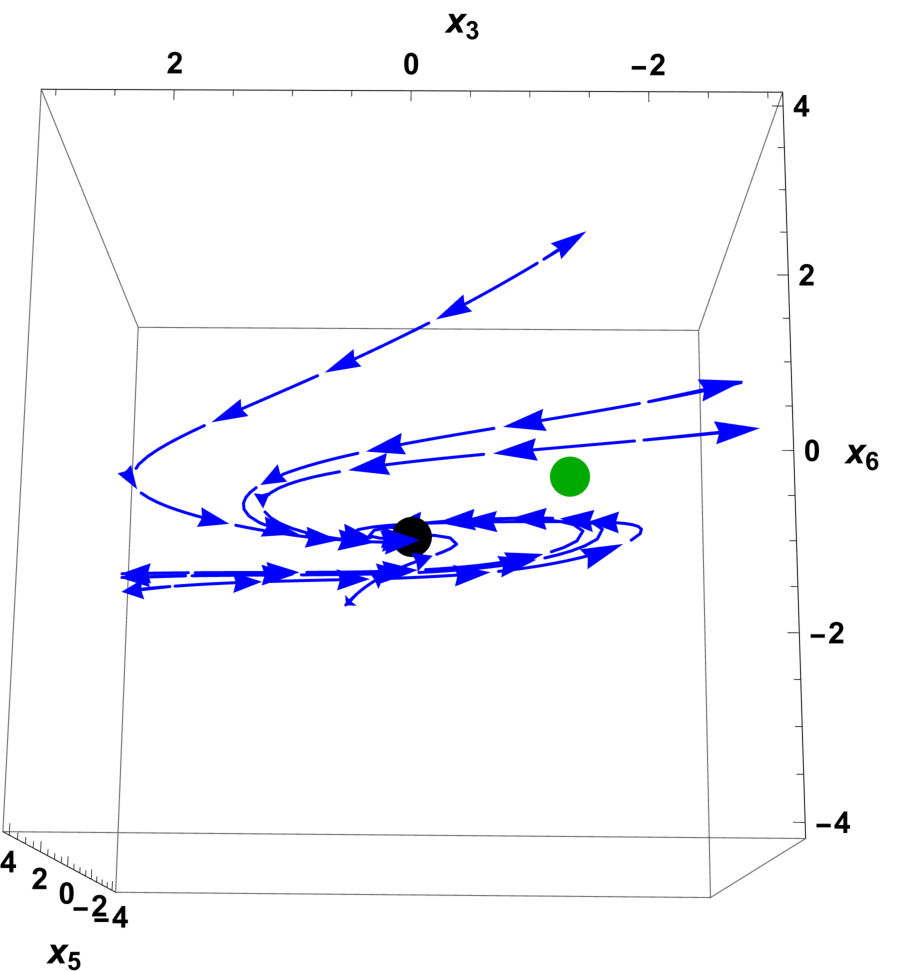}\hspace{2mm}
\caption{Cosmological scenarios obtained for connections type $III$ in the case $u=1, r=1/2$ and $v=2/3$. Here, point $P^{{\sf dS}}_{1}$ represents the early and late time accelerated expansion era, simultaneously. }\label{fig5}
\end{figure}

\begin{itemize}

\item $P^{{\sf dS}}_{1} \to P^{{\sf PLM}}_{1}/P^{{\sf PLM}}_{2}$ for $(v=-1/2, r=1/2)$: Since $P^{{\sf dS}}_{1}$ is a universal saddle point, independent of the values of $u$ and $v$, either $P^{{\sf PLM}}_{1}$ or $P^{{\sf PLM}}_{2}$ can connect to $P^{{\sf dS}}_{1}$ in phase space. Fig.~\ref{fig4} illustrates all possible scenarios constructed using connection type $III$. In this figure, the black, gray, green and darker green points represent $P^{{\sf dS}}_{1}$, $P^{{\sf dS}}_{2}$, $P^{{\sf PLM}}_{1}$ and $P^{{\sf PLM}}_{2}$, respectively. Note that the Universe evolves from $P^{{\sf dS}}_{1}$ to either $P^{{\sf PLM}}_{1}$ or $P^{{\sf PLM}}_{2}$, indicating that $P^{{\sf dS}}_{1}$ corresponds to an inflationary era. The upper-left and upper-right panels show the $P^{{\sf dS}}_{1}\to P^{{\sf PLM}}_{1}$ and $P^{{\sf dS}}_{1}\to P^{{\sf PLM}}_{2}$ transitions, respectively. Additionally, $P^{{\sf PLM}}_{2}$ exists for arbitrary values of $v$, but only for the particular case $r=1/2$. 

\item $P^{{\sf dS}}_{1} \to P^{{\sf PLM}}_{1} \& P^{{\sf PLM}}_{2}$: Models with $u=1$, $v=-1/2$ and $r=1/2$ fall into this category. A double {\sf PLM}-dominated era can be observed, as illustrated in the lower-left panel of Fig.~\ref{fig4}.

\item $P^{{\sf PLM}}_{1} \to P^{{\sf dS}}_{2}$: In this case, since $P^{{\sf dS}}_{2}$ is an attractor for all trajectories in the vicinity of $P^{{\sf PLM}}_{1}$ and exists for arbitrary values of $u$ and $v$, it connects to $P^{{\sf PLM}}_{1}$, which exists independently of $r$. Here, $P^{{\sf dS}}_{2}$ represents a late-time accelerated expansion, as shown in the lower-right panel of.~\ref{fig4}. In this category, the $P^{{\sf dS}}_{1} \to P^{{\sf PLM}}_{1} \to P^{{\sf dS}}_{2}$ scenario may also be reachable.

\item $P^{{\sf PLM}}_{2}\to P^{{\sf dS}}_{1}$ for $(v=2/3, r=1/2)$: As shown in Fig.~\ref{fig5}, point $P^{{\sf dS}}_{1}$ is a late time attractor for $v=2/3$. Even evolutions like $ P^{{\sf dS}}_{1} \to P^{{\sf PLM}}_{2} \to P^{{\sf dS}}_{1}$ is possible.

\item $P^{{\sf dS}}_{3} \to P^{{\sf PLM}}_{1}$:  $P^{{\sf dS}}_{3}$ is not stable, and therefore some trajectories originating near it will approach $P^{{\sf PLM}}_{1}$, making this scenario possible. A behavior similar to that shown in the upper-left panel of Fig.~\ref{fig4} is observed.

\item $P^{{\sf dS}}_{1}\to P^{{\sf Z}}(r=-\frac{3}{2})$: The point $P^{{\sf Z}}$, under the conditions mentioned in~(\ref{t3.17}), effectively represents a {\sf PLM}-dominated era when $r=-3/2$. Accordingly, the Universe evolves as described in the first item. It is clear that $P^{{\sf Z}}$ cannot describe a dark-energy era that connects to a matter-dominated epoch.
\end{itemize}

\subsection{$f(Q,C)=\alpha Q^{u}+\beta C^{v}$ with $u= v$}\label{type3.2}

Similar studies indicate that theories with $u= v$ lack a true fixed point describing a {\sf PLM}-dominated era. In fact, we identified a fixed point similar to~(\ref{t2.50}) which has the property $\Omega_{{\sf m}}|_{P^{{\sf PLM}}}=0$. Moreover, results like those discussed in section~\ref{type2.2.3} were not found. Finally, there exists a solution corresponding to {\sf SM} domination; however, it has $\Omega_{{\sf m}}|_{P^{{\sf SM}}}>2$.

\section{Concluding remarks}\label{conclusion}

The $f(Q,C)$ gravity framework, which integrates both non-metricity and boundary effects, has been formulated to overcome ambiguities and limitations inherent in General Relativity ({\sf GR}). Metricity is already implicit in the formal derivation of GR and other metric theories of gravity, while boundary effects correspond to the terms in the action that are typically discarded during the variation process leading to the field equations. By promoting both of these ingredients to dynamical relevance, $f(Q,C)$ gravity departs fundamentally from {\sf GR}, providing a novel foundation for exploring challenges across different domains of gravitational physics. In a cosmological set up, $f(Q,C)$ theories admit three types of affine connections that preserve spherical and translational symmetries~\cite{als/fQC}. Connection type $I$, corresponding to the vanishing case $\Gamma^{\alpha}_{\beta\gamma}=0$, leads to field equations that coincide with those of $f(T,B)$ gravity~\cite{baha23023}. Connection type $II$, characterized by the non-vanishing components given in 

\begin{align}
\Gamma^{t}{}_{tt}=\gamma(t)-3H(t),~~~~\Gamma^{i}{}_{it}=\gamma(t)=\Gamma^{i}{}_{ti},\nonumber
\end{align}

and connection type $III$, specified by the components

\begin{align}
\Gamma^{t}{}_{tt}=-H(t),~~~~\Gamma^{i}{}_{it}=\gamma(t)=\Gamma^{i}_{ti},\nonumber
\end{align}

with all others vanishing. Together, these three types encompass the admissible affine connections in $f(Q,C)$ cosmology. We analyzed the cosmological evolution by solving the equations of motion for specific functions of $\gamma$~\cite{shabani1:2025}. For $\gamma=\gamma_{0}a^{-3}$, giving $Q=-6H^{2},~C=6(3H^{2}+\dot{H})$, solutions with a Lagrangian formed from a linear combination of power-law functions of $Q$ and $C$ show that, in connection type $II$, the Universe evolves from an initial phantom phase to a final de Sitter era, with a {\sf PLM} fluid dominating the intermediate stage. For connection type $III$, using $\gamma=\gamma_{0}a^{-1}$ (yielding the same $Q$ and $C$), the Universe transitions from an inflationary phase, represented by a saddle de Sitter fixed point to a {\sf PLM}-dominated epoch.\\

In this work, we extend our previous analysis by considering the case $\gamma=\gamma_{0}a^{r}$ together with the functional form $f(Q,C)=\alpha Q^{u}+\beta C^{v}$, employing both type $II$ and type $III$ connections. The corresponding fixed point solutions are summarized in Tables~\ref{tab1} and~\ref{tab2} for the cases $u\neq v$ and $u=v$, respectively, within the framework of connection type $II$. The results obtained using connection type $III$ are presented in Table~\ref{tab3}. shows the solutions when connection type  is used. The main outcomes derived from connection type $II$, under the above assumptions, are outlined as follows.\\

\begin{itemize}
\item [\checkmark]  Connection type $II$, Case $u\neq v$:
  \(
  \left\{ \quad
    \begin{minipage}[c]{0.6\linewidth}
    \item A {\sf PLM}-dominated era (denoted by $P^{{\sf PLM}}$) arises for arbitrary values of $r$ when $u=1$ with either $v\approx 1$ or $v=-1/2$. In addition, another {\sf PLM}-dominated solution (labeled $P^{{\sf X}}$) emerges for $r=-3/2$, with $u=1,1/2$, and arbitrary $v$. Consequently, in the specific cases $u=1$, $v\approx 1$, $r=-3/2$, and $u=1$, $v=-1/2$, $r=-3/2$, two independent {\sf PLM} fixed points coexist simultaneously.

    \item A de Sitter fixed point exists for arbitrary values of $u$ and $r$, provided that $v \neq 1$.

    \item {\bf Results:} Different scenarios can be constructed featuring either one or two {\sf PLM} fixed points, within which the de Sitter fixed point corresponds to a late-time accelerated expansion era (see Fig.~\ref{fig1}). It should be emphasized that, in all scenarios, the de Sitter point corresponds to a late-time accelerated expansion when $v\approx 1$, whereas for $v=-1/2$ it instead characterizes an early inflationary epoch (see Fig.~\ref{fig1.1}).
    \end{minipage}
  \right.
  \)
\end{itemize}

\begin{itemize}
\item [\checkmark]  Connection type $II$, case $u=v$:
  \(
  \left\{ \quad
    \begin{minipage}[c]{0.6\linewidth}
    \item A single {\sf PLM}-dominated era (represented by $P^{\sf Y}_{1}$) arises for $v\gtrsim1$ with $r=3 (2-  v)/[(v+1) (2 v-3)]$.

    \item A de Sitter fixed point exists for $v\in \mathbb{Z}$ and arbitrary values of $r$, provided that $\alpha$ and $\beta$ satisfy the specific relation listed in Table~\ref{tab2}.

    \item There also exists a fixed point representing a stiff-matter-dominated era (denoted $P^{\sf SM}$). This solution arises for arbitrary values of $r$, $\alpha$, and $\beta$. When  $v\in (\frac{1}{2},1]$ the stiff-matter density parameter lies within the valid range.

\item {\bf Results:} Cosmological scenarios can be constructed either including or excluding $P^{\sf SM}$. In the case $u=v$ the de Sitter point can serve as both an early- and late-time accelerated expansion. The most intriguing scenario is
\[ P^{\sf dS}\to P^{\sf SM}\to P^{\sf PLM} \to P^{\sf dS},\]
which is supported by theories beyond the Standard Model~\cite{Pallis:2006,Visinelli:2018,Ghoshal:2022} (see lower-right panel in Fig.~\ref{fig2}).
    \end{minipage}
  \right.
  \)
\end{itemize}

Models with $u\approx 1$ and $v\approx 1$ indicate minimal deviations from {\sf GR}. In fact, in the trivial case $u=1$ and $v=1$, one recovers

\begin{align}
&3 \alpha  H^2=\kappa  \rho_{\sf m}\nonumber\\
&-2 \alpha  \dot{H}-3 \alpha  H^2=\kappa  p_{\sf m}\nonumber,
\end{align}

which correspond to the standard Friedmann equations with an effective gravitational coupling $\kappa_{f(Q,C)}=\kappa_{GR}/\alpha$. In this context, this class of $f(Q,C)$ theories may be considered particularly favorable. The properties of cosmological scenarios under connection type $III$ are analyzed as follows.

\begin{itemize}
\item [\checkmark]  Connection type $III$:
  \(
  \left\{ \quad
    \begin{minipage}[c]{0.6\linewidth}
    \item Here, the {\sf PLM}-dominated era is represented by two fixed points: $P^{\sf PLM}_{1}$, corresponding to $u=1, v=-1/2$, independent of $r$ and $P^{\sf PLM}_{2}$ with $u=1$ and $ r=1/2$, which is independent of $v$. In the particular model specified by $u=1$, $v=-1/2$ and $r=1/2$, both fixed points coexist simultaneously (see the lower-left panel of Fig.~\ref{fig4}).

    \item Three distinct critical points correspond to the de Sitter solution, denoted by $P^{\sf dS}_{1}$, $P^{\sf dS}_{2}$ and $P^{\sf dS}_{3}$. The point $P^{\sf dS}_{1}$ exists for arbitrary values of $u$ and $r$, provided that $v\neq 1$, and is stable only when $0<v<1$. The points $P^{\sf dS}_{2}$ and $P^{\sf dS}_{3}$ exist exclusively for $r=2$; among them, $P^{\sf dS}_{2}$ is always stable, while $P^{\sf dS}{3}$ behaves as a saddle point.

    \item {\bf Results:} The Universe can evolve from a de Sitter era to a {\sf PLM}-dominated phase, corresponding to either $P^{\sf PLM}_{1}$ or $P^{\sf PLM}_{2}$. It is also possible for the Universe to transition from a de Sitter era through two successive {\sf PLM}-dominated phases. Another scenario involves a temporary {\sf PLM} domination followed by a permanent de Sitter era (see Fig.~\ref{fig4}). Interestingly, an evolution from de Sitter to de Sitter, with an intervening {\sf PLM}-dominated phase, is also possible, in which $P^{\sf dS}_{1}$ represents both de Sitter epochs (see Fig.~\ref{fig5}). 
    \end{minipage}
  \right.
  \)
\end{itemize}

We conclude that $f(Q,C)$ gravity introduces novel cosmological scenarios, including the emergence of successive {\sf PLM}-dominated eras, which may interpreted as a baryonic matter-dominated phase following dark matter emergence; the occurrence of a {\sf SM}-dominated era, which contributes naturally to the cosmic evolution; de Sitter-to-de Sitter transitions realized solely through a de Sitter fixed point; and evolutionary paths encompassing either an early inflationary phase or late-time accelerated expansion.


\begin{thebibliography}{0}%
\makeatletter
\providecommand \@ifxundefined [1]{%
 \@ifx{#1\undefined}
}%
\providecommand \@ifnum [1]{%
 \ifnum #1\expandafter \@firstoftwo
 \else \expandafter \@secondoftwo
 \fi
}%
\providecommand \@ifx [1]{%
 \ifx #1\expandafter \@firstoftwo
 \else \expandafter \@secondoftwo
 \fi
}%
\providecommand \natexlab [1]{#1}%
\providecommand \enquote  [1]{``#1''}%
\providecommand \bibnamefont  [1]{#1}%
\providecommand \bibfnamefont [1]{#1}%
\providecommand \citenamefont [1]{#1}%
\providecommand \href@noop [0]{\@secondoftwo}%
\providecommand \href [0]{\begingroup \@sanitize@url \@href}%
\providecommand \@href[1]{\@@startlink{#1}\@@href}%
\providecommand \@@href[1]{\endgroup#1\@@endlink}%
\providecommand \@sanitize@url [0]{\catcode `\\12\catcode `\$12\catcode `\&12\catcode `\#12\catcode `\^12\catcode `\_12\catcode `\%12\relax}%
\providecommand \@@startlink[1]{}%
\providecommand \@@endlink[0]{}%
\providecommand \url  [0]{\begingroup\@sanitize@url \@url }%
\providecommand \@url [1]{\endgroup\@href {#1}{\urlprefix }}%
\providecommand \urlprefix  [0]{URL }%
\providecommand \Eprint [0]{\href }%
\providecommand \doibase [0]{https://doi.org/}%
\providecommand \selectlanguage [0]{\@gobble}%
\providecommand \bibinfo  [0]{\@secondoftwo}%
\providecommand \bibfield  [0]{\@secondoftwo}%
\providecommand \translation [1]{[#1]}%
\providecommand \BibitemOpen [0]{}%
\providecommand \bibitemStop [0]{}%
\providecommand \bibitemNoStop [0]{.\EOS\space}%
\providecommand \EOS [0]{\spacefactor3000\relax}%
\providecommand \BibitemShut  [1]{\csname bibitem#1\endcsname}%
\let\auto@bib@innerbib\@empty
\end{thebibliography}%


\begin{thebibliography}{99}


\bibitem{heisenberg}
L.~Heisenberg, 
``A systematic approach to generalisations of General Relativity and their cosmological implications,'' 
Phys.\ Rept.\ \textbf{796} (2019) 1--113.

\bibitem{laur} 
L.~Järv, M.~Rünkla, M.~Saal, and O.~Vilson, 
``Nonmetricity formulation of general relativity and its scalar-tensor extension,'' 
Phys.\ Rev.\ D \textbf{97} (2018) 124025.

\bibitem{ghulamqphi1} 
G.~Murtaza, A.~De, A.~Paliathanasis, and T.H.~Loo, 
``Can an Extra Degree of Freedom in Scalar-Tensor Non-Metricity Gravity Account for the Evolution of the Universe?,'' 
Class.\ Quant.\ Grav.\ (2025) in press, 
doi:10.1088/1361-6382/ae0404 [arXiv:2506.17099 [gr-qc]].

\bibitem{ghulamqphi2} 
G.~Murtaza, A.~De, A.~Paliathanasis, 
``A novel non-metricity extension of scalar-tensor gravity in spatially curved spacetime,'' 
arXiv:2509.00569 [gr-qc].

\bibitem{coincident} 
J.~B.~Jiménez, L.~Heisenberg and T.~Koivisto, 
``Coincident General Relativity,'' 
Phys.\ Rev.\ D \textbf{98} (2018) 044048.

\bibitem{fqsurvey}
L.~Heisenberg, 
``Review on $f(Q)$ gravity,'' 
Phys.\ Rept.\ \textbf{1066} (2024) 1--78.

\bibitem{ghulamfqbi}
G.~Murtaza, S.~Chakraborty, and A.~De, 
``A generic dynamical system formulation for Bianchi-I cosmology with isotropic fluid in $f(Q)$ gravity,'' 
JCAP \textbf{08} (2025) 093.

\bibitem{als/fQC}  
A.~De, T.~H.~Loo and E.~N.~Saridakis, 
``Non-metricity with boundary terms: $f(Q,C)$ gravity and cosmology,''     
JCAP \textbf{03} (2024) 050 [arXiv:2308.00652 [gr-qc]].

\bibitem{Capozziello2023a} 
S.~Capozziello, V.~De Falco, and C.~Ferrara, 
``The role of the boundary term in $f(Q,B)$ symmetric teleparallel gravity,'' 
Eur.\ Phys.\ J.\ C \textbf{83} (2023) 915.

\bibitem{shabani1:2025}
H.~Shabani, A.~De, and T.H.~Loo, 
``The cosmological significance of boundary term in non-metricity gravity,'' 
Nucl.\ Phys.\ B \textbf{1017} (2025) 116965.

\bibitem{ghulamannals} 
G.~Murtaza, A.~De, T.H.~Loo, Y.~K.~Goh, and H.H.~Liew, 
``Phase-space analysis of an anisotropic universe in $f(Q,C)$ gravity,'' 
Annals Phys.\ \textbf{480} (2025) 170086.

\bibitem{samaddar2025} 
A.~Samaddar and S.S.~Singh, 
``Cosmological dynamics and thermodynamic behavior in $f(Q,C)$ gravity: An analytical and observational approach,'' 
Phys.\ Dark Univ.\ \textbf{42} (2025) 101792.

\bibitem{Sadatian2024} 
S.D.~Sadatian and S.M.R.~Hosseini, 
``Cosmological inflation in the modified gravity model $f(Q,C)$,'' 
Phys.\ Dark Univ.\ \textbf{41} (2024) 101737.

\bibitem{Sharif2025}  
M.~Sharif, M.Z.~Gul, and M.H.~Shahid, 
``Cosmological implications and ghost dark energy model in $f(Q,C)$ gravity,'' 
High Energy Density Phys.\ \textbf{47} (2025) 101185.

\bibitem{Alruwaili2025} 
A.D.~Alruwaili, N.~Azhar, A.~Jawad, and A.B.~Albarrak, 
``Probing gravitational baryogenesis phenomenon in specific modified gravity,'' 
Phys.\ Dark Univ.\ \textbf{42} (2025) 101842.

\bibitem{ppaliathanasis2024}
A.~Paliathanasis, 
``Attractors in $f(Q,B)$ gravity,'' 
Phys.\ Dark Univ.\ \textbf{45} (2024) 101519.

\bibitem{Muhammad2024} 
A.~Samaddar, S.S.~Singh, S.~Muhammad, and E.E.~Zotos, 
``Behaviours of rip cosmological models in $f(Q,C)$ gravity,'' 
Nucl.\ Phys.\ B \textbf{1006} (2024) 116643.

\bibitem{BH}
J.T.S.S.~Junior, F.S.N.~Lobo, and M.E.~Rodrigues, 
``Black holes and regular black holes in coincident $f(Q,B_Q)$ gravity coupled to nonlinear electrodynamics,'' 
Eur.\ Phys.\ J.\ C \textbf{84} (2024) 332.

\bibitem{bmishra}
S.V.~Lohakare and B.~Mishra, 
``Stability of $f(Q,B)$ Gravity via Dynamical System Approach: A Comprehensive Bayesian Statistical Analysis,'' 
Astrophys.\ J.\ \textbf{978} (2025) 26.

\bibitem{recon}
G.N.~Gadbail, A.~De, and P.K.~Sahoo, 
``Cosmological reconstruction and $\Lambda$CDM universe in $f(Q,C)$ gravity,'' 
Eur.\ Phys.\ J.\ C \textbf{83} (2023) 1099.

\bibitem{ganeshpert} 
G.~Subramaniam, A.~De, T.H.~Loo, and Y.K.~Goh, 
``Scalar perturbation and density contrast evolution in $f(Q,C)$ gravity,'' 
arXiv:2412.05382 [gr-qc].

\bibitem{baha23023}  
S.~Bahamonde et al., 
``Teleparallel gravity: from theory to cosmology,'' 
Rept.\ Prog.\ Phys.\ \textbf{86} (2023) 026901.

\bibitem{avikand}
A.~De and A.~Paliathanasis, 
``Exact cosmological solutions in non-coincidence $f(Q)$-theory,'' 
arXiv:2509.14174 [gr-qc].

\bibitem{Paliathanasis:2023nkb}
A.~Paliathanasis, 
``Dynamical system analysis in $f(Q)$ gravity,'' 
Phys.\ Dark Univ.\ \textbf{41} (2023) 101255 [arXiv:2304.04219 [gr-qc]].

\bibitem{shabanifQ}
H.~Shabani, A.~De, and T.H.~Loo, 
``Dynamics in $f(Q)$ cosmology,'' 
Eur.\ Phys.\ J.\ C \textbf{83} (2023) 535.

\bibitem{Narawade:2024pxb}
S.A.~Narawade, S.V.~Lohakare, and B.~Mishra, 
``Cosmological implications of $f(Q)$ gravity,'' 
Annals Phys.\ \textbf{474} (2025) 169913 [arXiv:2408.12846 [gr-qc]].

\bibitem{ghulamfQ}
G.~Murtaza, S.~Chakraborty, and A.~De, 
``Generic dynamical system formulation for $f(Q)$ gravity,'' 
JCAP \textbf{08} (2025) 093.

\bibitem{saikatfQ}
J.~Dutta, W.~Khyllep, S.~Chakraborty, D.~Gregoris, and K.~Karwan, 
``Cosmological dynamics in $f(Q)$ gravity,'' 
arXiv:2508.09530 [gr-qc] (2025).

\bibitem{rajadata}
S.~Pradhan, R.~Solanki, and P.K.~Sahoo, 
``Observational data analysis in $f(Q)$ gravity,'' 
J.\ High Energy Astrophys.\ \textbf{43} (2024) 258--267.

\bibitem{saridakisdata}
Y.~Yang, X.~Ren, B.~Wang, Y.F.~Cai, and E.N.~Saridakis, 
``Observational constraints on $f(Q)$ gravity,'' 
Mon.\ Not.\ Roy.\ Astron.\ Soc.\ \textbf{533} (2024) 2232--2241.

\bibitem{saikatfQlcdm}
S.~Chakraborty, J.~Dutta, D.~Gregoris, K.~Karwan, and W.~Khyllep, 
``$\Lambda$CDM reconstruction in $f(Q)$ cosmology,'' 
JCAP \textbf{05} (2025) 098.

\bibitem{an01}
A.~Paliathanasis, 
``Testing non-coincident $f(Q)$-gravity with DESI DR2 BAO and GRBs,'' 
Phys.\ Dark Univ.\ \textbf{49} (2025) 101993.

\bibitem{Amendola:2007}
L.~Amendola, R.~Gannouji, D.~Polarski, and S.~Tsujikawa,
``Conditions for the cosmological viability of $f(R)$ dark energy models,'' 
Phys.\ Rev.\ D \textbf{75} (2007) 083504.

\bibitem{FLRW/connection} 
N.~Dimakis, A.~Paliathanasis, M.~Roumeliotis, and T.~Christodoulakis, 
``FLRW solutions in $f(Q)$ theory: the effect of using different connections,'' 
Phys.\ Rev.\ D \textbf{106} (2022) 043509.

\bibitem{ad/bianchi}
A.~De and T.H.~Loo, 
``On the viability of $f(Q)$ gravity models,'' 
Class.\ Quant.\ Grav.\ \textbf{40} (2023) 115007.


\bibitem{Pallis:2006}
C.~Pallis, 
``Kination-dominated reheating and cold dark matter abundance,'' 
Nucl.\ Phys.\ B \textbf{751} (2006) 129.

\bibitem{Visinelli:2018}
L.~Visinelli, 
``(Non-)Thermal Production of WIMPs during Kination,'' 
Symmetry \textbf{10} (2018) 546.

\bibitem{Ghoshal:2022}
A.~Ghoshal, L.~Heurtier, and A.~Paul, 
``Signatures of non-thermal dark matter with kination and early matter domination: Gravitational waves versus laboratory searches,'' 
JHEP \textbf{12} (2022) 105.

\end{thebibliography}
\end{document}